\documentclass{iopjournal}

\usepackage[T1]{fontenc}
\usepackage[utf8]{inputenc}
\usepackage{amsmath,amssymb,amsfonts}
\usepackage{mathrsfs}
\usepackage{booktabs}
\usepackage{tabularx}
\usepackage{multirow}
\usepackage[numbers,sort&compress]{natbib}
\usepackage{url}
\usepackage{seqsplit}
\usepackage{placeins}
\begin{document}

\articletype{Paper}

\title{Derivative-consistent finite-domain inverse reconstruction of an
$f(R)$ curvature sector for a Morris--Thorne geometry}

\author{Murat Metehan T\"urko\u{g}lu}

\affil{Istanbul Gelisim University, Department of Aeronautical Engineering,\ Cihangir Mah. \c{S}ehit Jandarma Komando Er Hakan \"Oner Sk. No:1, Istanbul, T\"urkiye}

\email{mmturkoglu@gelisim.edu.tr}


\begin{abstract}
We formulate the finite-domain reconstruction of a prescribed
Morris--Thorne geometry as a constrained inverse problem for a
derivative-consistent $f(R)$ curvature sector. Given metric functions
whose throat and local flare-out conditions are imposed analytically
and whose redshift profile is bounded, the resulting curvature
trajectory $R(r)$ is treated as the sampled input on which an
admissible curvature function is reconstructed. The formulation does
not require a numerical inverse $r=r(R)$ and therefore remains
well-defined when the sampled curvature trajectory is non-monotonic.
Instead of fitting $f(R)$, $f_R(R)$, and $f_{RR}(R)$ independently, a
positive analytic generator is assigned to $f_{RR}(R)$ and integrated
successively to obtain $f_R(R)$ and $f(R)$. Hence
$df/dR=f_R$ and $df_R/dR=f_{RR}$ hold by construction, eliminating
derivative drift between independently approximated curvature
quantities. The method is evaluated on one frozen reconstruction with
10,019 radial nodes over $r\in[1,10^{5}]$. The sampled trajectory spans
$-1.1317\times10^{-9}\leq R\leq1.999999998$ and contains a turning
point, while the reconstructed branch retains $f_R>0$ and $f_{RR}>0$
at every audited node. On the same frozen object, all reported
reconstructed-source energy-condition combinations remain positive,
the archived source-side closure has maximum norm
$5.3246\times10^{-5}$, and direct arithmetic recomputation agrees to
within $8.33\times10^{-17}$. These results establish an auditable
finite-domain inverse-reconstruction benchmark in which analytic
constraints, curvature recovery, and post-selection diagnostics remain
separated without result-level refitting. The source representation is
phenomenological; no unique microscopic matter Lagrangian or
curvature--matter coupling is inferred. Accordingly, the benchmark
establishes finite-domain numerical consistency and admissibility
diagnostics rather than global existence, dynamical stability,
exterior matching, or observer-dependent safe traversal.
\end{abstract}

\section{Introduction}\label{sec:introduction}

Inverse problems seek unknown parameters, functions, or latent physical
quantities from quantities that are more directly prescribed, measured,
or constrained. In such problems, reproducing a target or obtaining a
small numerical residual is not sufficient by itself: the reconstructed
object must also belong to an admissible class and preserve the
structural relations required by the underlying model
\cite{Tarantola2005,EnglHankeNeubauer1996}. This distinction becomes
particularly important in functional inverse problems, where several
numerical representations may approximate the same sampled information
while differing in differentiability, sign properties, conditioning, or
internal compatibility.

The problem considered here is a deterministic, geometry-conditioned
functional inverse reconstruction. A prescribed static spacetime
geometry determines a sampled Ricci-scalar trajectory
\[
R=R(r),
\]
and the inverse task is to reconstruct an admissible curvature function
$f(R)$ on the sampled curvature interval. The reconstructed hierarchy
must satisfy
\[
\frac{df}{dR}=f_R,
\qquad
\frac{df_R}{dR}=f_{RR},
\]
while remaining compatible with selected finite-domain curvature and
source-side diagnostics. The term \emph{inverse reconstruction} is used
in this specific sense throughout the paper. We do not claim recovery
from noisy observational data, uniqueness of the reconstructed
functional model, generic Hadamard well-posedness, or a regularization
theorem for an unrestricted inverse problem.

A Morris--Thorne geometry provides a stringent physical test case for
this reconstruction. A static and spherically symmetric wormhole must
contain a throat, satisfy the local flare-out condition, avoid a horizon
through a finite redshift function, and remain compatible with further
geometric constraints relevant to traversability
\cite{MorrisThorne1988,Visser1995}. In Einstein gravity, the throat
conditions are normally associated with violation of the null energy
condition by the matter sector, giving rise to the familiar
exotic-matter obstruction. Quantum inequalities further constrain the
magnitude and duration of negative-energy configurations that can be
maintained in semiclassical settings \cite{FordRoman1995}. The
Morris--Thorne geometry is therefore useful here not merely as an
application of modified gravity, but as a highly constrained testbed in
which geometric admissibility, curvature reconstruction, source-side
properties, and numerical consistency must remain distinguishable.

Modified-gravity theories enlarge the possible relation between
spacetime curvature and source terms. In $f(R)$ gravity the
Einstein--Hilbert curvature dependence is replaced by a nonlinear
function of the Ricci scalar, whereas nonminimal
curvature--matter-coupling theories introduce additional
curvature-dependent source interactions
\cite{BertolamiEtAl2008,HarkoLobo2010}. Analytical wormhole studies
have shown that these modifications can redistribute the effective
throat-support balance and, under particular model assumptions, allow
selected matter variables to satisfy standard energy-condition
inequalities
\cite{GarciaLobo2010a,GarciaLobo2010b,
DeBenedictisHorvat2011,PavlovicSossich2014}. Related constructions in
$f(R)$, $f(R,L_m)$, $f(R,T)$, and other matter--geometry extensions
nevertheless demonstrate that such conclusions depend strongly on the
chosen geometry, matter Lagrangian, equation of state, coupling
prescription, and boundary assumptions
\cite{SamantaGodani2019,GolchinMehdizadeh2019,
GhoshMitraChakraborty2021,VenkateshaEtAl2023,
RosaKull2022,BanerjeeTangphatiPradhan2023,
RosaGaniyevaLobo2023,GaniyevaRosaLobo2025}.
Consequently, satisfaction of an energy-condition inequality in one
reconstructed representation cannot by itself establish that exotic
support has been universally eliminated from a fully specified theory.

From the inverse-problem perspective, a distinct numerical difficulty
arises before such physical interpretation is attempted. Suppose that
$f(R)$, $f_R(R)$, and $f_{RR}(R)$ are approximated independently on a
finite grid. Even if all three arrays separately reproduce selected
targets with small residuals, nothing guarantees that they form the
derivative hierarchy of one common function. In general,
\[
f_R^{\mathrm{num}}(R)
\neq
\frac{d f^{\mathrm{num}}}{dR},
\qquad
f_{RR}^{\mathrm{num}}(R)
\neq
\frac{d f_R^{\mathrm{num}}}{dR}.
\]
The reconstruction may therefore satisfy result-level diagnostics while
being structurally inconsistent. This is an admissibility problem that
cannot be resolved simply by reporting a small optimization loss.

A second difficulty arises when the prescribed geometry generates a
non-monotonic curvature trajectory. Any procedure that first attempts
to invert $R(r)$ globally and construct $r=r(R)$ may then encounter
multiple radial locations associated with the same curvature value.
Such an inversion is unnecessary if the reconstructed curvature sector
is represented directly as a single-valued function of $R$. This
observation motivates the two central design choices of the present
method: derivative consistency is imposed through the functional
parameterization itself, and the reconstruction is evaluated directly
on the sampled curvature trajectory without constructing a global
inverse map $r(R)$.

Specifically, instead of fitting $f(R)$, $f_R(R)$, and $f_{RR}(R)$ as
three nominally related numerical objects, we parameterize a positive
analytic generator for $f_{RR}(R)$ and obtain the lower-order functions
by successive integration,
\[
f_{RR}(R)
\longrightarrow
f_R(R)
\longrightarrow
f(R).
\]
The identities
\[
\frac{df_R}{dR}=f_{RR},
\qquad
\frac{df}{dR}=f_R
\]
therefore hold by construction. The numerical search is restricted to a
smaller admissible class in which derivative mismatch is excluded
structurally rather than penalized after it occurs. This construction
does not solve generic nonuniqueness of the inverse problem; rather, it
removes one class of internally inconsistent candidate solutions from
the admissible reconstruction space.

Physics-informed learning provides a useful computational setting for
constructing constrained functional representations
\cite{RaissiEtAl2019}. Related neural approaches have been applied to
Einstein equations, quasinormal-mode calculations, and modified-gravity
problems
\cite{LiLiPang2023,LunaEtAl2024,NcubeHarmsenCornell2021,
CornellNcubeHarmsen2022,AitHaddou2023}. However, physics-informed
optimization can suffer from loss imbalance, gradient pathologies,
sampling sensitivity, stiffness, and optimizer-dependent failure modes
\cite{WangTengPerdikaris2020,CuomoEtAl2022}. For this reason, the
present evidential strategy does not identify training loss with
physical validity. Analytic geometric constraints and derivative
identities are separated from the later numerical diagnostics, and the
selected reconstruction is frozen before the final validation layer is
evaluated.

The methodological problem addressed here can therefore be stated more
precisely. We seek a finite-domain reconstruction that simultaneously
prevents three distinct failure modes: violation of defining geometric
conditions by an otherwise flexible numerical representation,
derivative inconsistency among $f$, $f_R$, and $f_{RR}$, and
post-selection adjustment of different diagnostics on mutually
inconsistent numerical objects. The first is controlled by analytic
parameterization of the Morris--Thorne geometry; the second by the
integrated curvature hierarchy; and the third by evaluating the final
diagnostics on one frozen reconstruction.

The resulting proof-of-principle is evaluated on
$N=10019$ radial nodes over the throat-scaled interval
$r\in[1,10^5]$. The corresponding sampled Ricci-scalar trajectory spans
\[
-1.1317148307\times10^{-9}
\leq R \leq
1.9999999979
\]
and is non-monotonic, with one detected turning point. No numerical
inverse $r=r(R)$ is used. On the same frozen object, the reconstructed
curvature branch satisfies $f_R>0$ and $f_{RR}>0$ at every audited
node. The reconstructed-source energy-condition combinations reported
in the study remain positive on the finite grid, while the archived
source-side closure has maximum norm
$5.324612936468367\times10^{-5}$ and a relative maximum-norm reduction
of approximately $99.992850\%$. Direct arithmetic recomputation of the
stored closure channels agrees to within
$8.33\times10^{-17}$.

These numerical results serve two purposes. First, they demonstrate that
the analytic admissibility constraints and derivative-consistent
curvature reconstruction can coexist on a nontrivial finite-domain
geometry. Second, they establish a reproducible frozen object on which
geometric, curvature-sector, source-side, tidal, and numerical-closure
diagnostics can be compared without result-level refitting. The
methodological contribution is therefore not a claim that machine
learning has discovered a unique wormhole-supporting theory. It is a
constrained inverse-reconstruction strategy in which structural
consistency is imposed before the numerical evidence is interpreted.

The underlying numerical principle is not specific to wormholes.
Whenever an inverse problem contains unknown functions connected by
exact differential identities, representing one member of the
derivative hierarchy and obtaining the remaining members through
integration can restrict the search to structurally admissible
solutions. The Morris--Thorne geometry provides the physical stress
test in the present work, whereas the derivative-consistent
reconstruction principle can in principle be transferred to other
functional inverse problems whose unknown quantities are linked by
known differential relations. Establishing such transferability in
other physical systems is left for future work.

Finally, the physical claim boundary is explicit. The quantities
$\rho$, $p_r$, and $p_t$ used below are phenomenologically
reconstructed source variables rather than a stress--energy tensor
derived from a uniquely specified microscopic matter Lagrangian. The
archived source-side closure is likewise a numerical closure diagnostic,
not an independent covariant re-solution of a uniquely specified
nonminimally coupled $f(R)$ theory. The present reconstruction therefore
does not establish a unique curvature--matter coupling, global
existence or uniqueness, exterior matching, perturbative stability,
affine-complete averaged-null-energy satisfaction, or
observer-dependent safe traversal. Section~2 formalizes this evidence
hierarchy before the geometric and curvature-sector constructions are
introduced.

\section{Reference theory, instantiated inverse reconstruction, and claim boundary}

The inverse reconstruction introduced in Sec.~1 is motivated by
modified-curvature and curvature--matter-coupled gravity, but the
reference gravitational theory must be distinguished from the
numerical object that is actually reconstructed. This distinction is
particularly important for the present inverse-problem formulation:
the generic curvature--matter-coupling equations define a theoretical
context and identify quantities that would be required by a fully
specified physical model, whereas the finite-domain reconstruction
uses only those quantities that are explicitly instantiated and
archived. A representative nonminimal curvature--matter coupling
therefore provides a useful reference for understanding how
curvature-dependent terms can alter the balance between geometry and
explicit matter sources
\cite{BertolamiEtAl2008,HarkoLobo2010}, and such couplings have been
used directly in analytical wormhole constructions
\cite{GarciaLobo2010a,GarciaLobo2010b}. The present calculation,
however, does not identify a unique microscopic matter Lagrangian, a
unique nonminimal coupling strength, or a unique coupling function
whose complete covariant field equations are independently re-solved.
The reference equations below consequently delimit the theoretical
model class and the associated claim boundary; they are not presented
as a more specific numerical forward model than has actually been
instantiated.

\subsection{Reference curvature--matter-coupling framework}

A representative nonminimally coupled $f(R)$ action may be written
in the form
\begin{equation}
S_{\rm ref}
=
\int d^4x\,\sqrt{-g}
\left[
\frac{f(R)}{2\kappa}
+
A(R)\mathcal L_m
\right],
\label{eq:reference_nmc_action}
\end{equation}
where $\kappa$ denotes the gravitational coupling,
$\mathcal L_m$ is a matter Lagrangian, and
\begin{equation}
A(R)=1+\lambda h(R)
\label{eq:reference_coupling}
\end{equation}
is a curvature-dependent coupling function
\cite{BertolamiEtAl2008,HarkoLobo2010}. Defining
\begin{equation}
A_R(R)
\equiv
\frac{dA}{dR}
=
\lambda h_R(R),
\qquad
h_R(R)
\equiv
\frac{dh}{dR},
\end{equation}
the curvature derivative entering the metric variation of the
reference theory is
\begin{equation}
F_{\rm NMC}
\equiv
f_R
+
2\kappa A_R\mathcal L_m,
\qquad
f_R\equiv\frac{df}{dR}.
\label{eq:F_NMC_reference}
\end{equation}

For the reference action in Eq.~\eqref{eq:reference_nmc_action},
the metric field equation may be written schematically as
\begin{equation}
\begin{aligned}
F_{\rm NMC}R_{\mu\nu}
-
\frac{1}{2}f(R)g_{\mu\nu}
&+
\left(g_{\mu\nu}\Box-\nabla_\mu\nabla_\nu\right)F_{\rm NMC}
\\
&=
\kappa A(R)T_{\mu\nu},
\end{aligned}
\label{eq:reference_nmc_field_equation}
\end{equation}
where
\begin{equation}
T_{\mu\nu}
=
-\frac{2}{\sqrt{-g}}
\frac{\delta\!\left(\sqrt{-g}\mathcal L_m\right)}
{\delta g^{\mu\nu}}.
\end{equation}

Equations~\eqref{eq:reference_nmc_action}--%
\eqref{eq:reference_nmc_field_equation} define the theoretical
reference structure that motivates the separation between curvature
and source-side quantities. They are not the fully instantiated
numerical equations of the frozen inverse reconstruction. In
particular, the numerical archive does not assign a unique physical
value of $\lambda$, a unique function $h(R)$, or a unique microscopic
$\mathcal L_m$ from which a complete candidate-specific
$F_{\rm NMC}(r)$ could be evaluated. Consequently, the reference
framework is used to state what additional information a complete
physical model would require, rather than to infer quantities that
are absent from the reconstruction.

This distinction is important in the present context. Analytical
nonminimal-coupling wormhole models demonstrate that curvature and
coupling contributions can alter the effective throat-support
balance under explicitly specified model assumptions
\cite{GarciaLobo2010a,GarciaLobo2010b}. The present work uses that
literature as theoretical motivation, but does not infer that the
same microscopic interpretation follows automatically from a
phenomenologically reconstructed source-side closure.

\subsection{Instantiated finite-domain inverse-reconstruction object}

The computational problem actually solved in this work is defined by
the quantities that are explicitly reconstructed and archived on the
finite radial grid. Once the geometry determines the sampled
trajectory $R(r)$, the curvature-sector inverse reconstruction and all
subsequent diagnostics are evaluated on this common finite-domain
object. Denoting the radial grid by
\begin{equation}
\mathcal D_r
=
\{r_i\}_{i=1}^{N},
\qquad
r_i\in[1,10^5],
\qquad
N=10019,
\end{equation}
the load-bearing frozen reconstruction may be represented as
\begin{equation}
\begin{aligned}
\mathcal C_{\rm frz}
=\{&r_i,b_i,\Phi_i,H_i,R_i,f_i,f_{R,i},f_{RR,i},
\\
&\rho_i,p_{r,i},p_{t,i},\mathcal T_i,
Q^{\rm base}_{c,i},Q^{\rm src}_{c,i}\}.
\end{aligned}
\label{eq:frozen_reconstruction_object}
\end{equation}
with
\begin{equation}
c\in
\{tt,rr,\theta\theta,\mathrm{trace}\}.
\end{equation}
Here
\begin{equation}
H_i
=
1-\frac{b_i}{r_i},
\qquad
R_i=R(r_i),
\end{equation}
and
\begin{equation}
f_i=f(R_i),
\qquad
f_{R,i}=f_R(R_i),
\qquad
f_{RR,i}=f_{RR}(R_i).
\end{equation}

The metric and curvature-sector quantities are reconstructed
explicitly and are described in Secs.~3 and 4. The quantities
$\rho_i$, $p_{r,i}$, and $p_{t,i}$ are retained as
\emph{reconstructed source variables}. They are used to form the
finite-domain energy-condition and radial null-energy diagnostics,
but they are not identified with a stress--energy tensor derived
from a unique microscopic matter Lagrangian.

Likewise, the archived channel quantities
$Q^{\rm base}_{c,i}$ and $Q^{\rm src}_{c,i}$ are treated as the two
sides of the reconstructed source-side closure representation.
Their difference defines the channel-wise closure diagnostic
\begin{equation}
\mathcal C_c(r_i)
=
Q^{\rm base}_{c,i}
-
Q^{\rm src}_{c,i},
\label{eq:closure_diagnostic}
\end{equation}
and the corresponding finite-domain maximum norm is
\begin{equation}
\|\mathcal C\|_{\infty,\mathcal D_r}
=
\max_c\max_i
\left|
\mathcal C_c(r_i)
\right|.
\label{eq:closure_maxnorm_definition}
\end{equation}

Equation~\eqref{eq:closure_diagnostic} is an internal numerical
closure diagnostic of the frozen reconstruction. It is deliberately
not identified with an independent covariant recomputation of
Eq.~\eqref{eq:reference_nmc_field_equation} for a uniquely specified
set
$\{\lambda,h(R),\mathcal L_m\}$.
This terminology is used throughout the remainder of the paper.

\subsection{GR reference and finite-domain curvature admissibility diagnostics}

The Einstein--Hilbert curvature sector provides a useful reference
point for interpreting the reconstructed $f(R)$ branch. In the
presence of a cosmological constant, the corresponding curvature
function is
\begin{equation}
f_{\rm GR}(R)=R-2\Lambda,
\qquad
f_{R,\rm GR}=1,
\qquad
f_{RR,\rm GR}=0.
\label{eq:GR_reference}
\end{equation}
A convenient curvature-sector departure measure is therefore
\begin{equation}
\Delta_{\rm GR}(R)
\equiv
f_R(R)-1.
\label{eq:Delta_GR_reference}
\end{equation}
This quantity characterizes only the reconstructed curvature sector;
it does not quantify a candidate-specific nonminimal matter
coupling.

For a fully specified model of the reference form
in Eq.~\eqref{eq:reference_nmc_action}, one could formally define
\begin{equation}
\Delta_{\rm NMC}^{\rm ref}(R,r)
=
2\kappa\lambda h_R(R)\mathcal L_m(r),
\label{eq:Delta_NMC_formal}
\end{equation}
so that
\begin{equation}
F_{\rm NMC}
=
1
+
\Delta_{\rm GR}
+
\Delta_{\rm NMC}^{\rm ref}.
\end{equation}
No numerical value or radial profile of
$\Delta_{\rm NMC}^{\rm ref}$ is reported in the present work,
because the frozen reconstruction does not uniquely determine
$\lambda$, $h_R(R)$, and $\mathcal L_m(r)$. Consequently,
$\Delta_{\rm NMC}^{\rm ref}$ is not used as a validation gate or as
evidence for the physical support mechanism of the reconstructed
candidate.

The admissibility diagnostics used for the reconstructed curvature
sector are therefore restricted to quantities that are directly
available from the instantiated $f(R)$ branch. In particular,
\begin{equation}
f_R(R_i)>0,
\qquad
f_{RR}(R_i)>0
\end{equation}
are audited on the sampled curvature trajectory. These conditions
are commonly used as local viability-oriented diagnostics in
$f(R)$ gravity and in $f(R)$ wormhole studies
\cite{DeBenedictisHorvat2011,PavlovicSossich2014,
NaluiBhattacharya2025}. In the present work they are interpreted
only as finite-domain sign diagnostics of the reconstructed
curvature branch. They are not promoted to a global stability
theorem, and they do not determine the sign of a nonminimal
prefactor that has not been uniquely instantiated.

\subsection{Evidence hierarchy and claim boundary}

The admissible inverse-reconstruction object and the quantities
belonging only to the reference physical theory therefore occupy
different evidential levels. Table~\ref{tab:theory_numeric_boundary}
makes this distinction explicit so that the generic
curvature--matter-coupling motivation is not confused with the
quantities that are actually reconstructed.

\begin{table*}[t]
\centering
\caption{
Separation between the theoretical reference framework and the
quantities explicitly instantiated in the finite-domain numerical
reconstruction. ``Not uniquely specified'' means that the quantity
is not assigned a candidate-specific numerical form and is therefore
not used to support a stronger physical claim.
}
\label{tab:theory_numeric_boundary}
\small
\begin{tabularx}{\textwidth}{@{}>{\raggedright\arraybackslash}p{2.8cm}>{\raggedright\arraybackslash}p{3.2cm}>{\raggedright\arraybackslash}X@{}}
\hline
\textbf{Quantity or sector} &
\textbf{Status in this work} &
\textbf{Role and claim boundary} \\
\hline

Morris--Thorne geometry
&
Explicitly reconstructed
&
$b(r)$ and $\Phi(r)$ define the finite-domain geometry and are
subject to the throat, flare-out, redshift, and no-horizon
diagnostics. \\

$f(R)$ curvature sector
&
Explicitly reconstructed
&
$f(R)$, $f_R(R)$, and $f_{RR}(R)$ are generated as one
derivative-consistent curvature-sector object on the sampled
curvature interval. \\

$\rho$, $p_r$, $p_t$
&
Phenomenologically reconstructed
&
Used as reconstructed source variables for finite-domain
energy-condition and radial null-energy diagnostics; not derived
from a unique microscopic matter Lagrangian. \\

$\lambda$
&
Not uniquely specified
&
No candidate-specific nonminimal coupling strength is inferred or
used as a numerical validation parameter. \\

$h(R)$ and $h_R(R)$
&
Not uniquely specified
&
The reference coupling function motivates the theoretical
source--curvature separation but is not claimed as a uniquely
reconstructed physical function. \\

$\mathcal L_m$
&
Not uniquely specified
&
No unique microscopic matter Lagrangian is derived for the
reconstructed source variables. \\

$F_{\rm NMC}$
&
Reference-theory quantity
&
The formal combination
$f_R+2\kappa A_R\mathcal L_m$ belongs to the generic NMC framework;
a complete candidate-specific profile is not used as an independently
recomputed numerical field-equation operator. \\

Archived closure channels
&
Explicitly available
&
$Q_c^{\rm base}-Q_c^{\rm src}$ defines an internal source-side
closure diagnostic of the frozen reconstruction; it is not presented
as an independent covariant re-solution of a uniquely specified NMC
theory. \\

\hline
\end{tabularx}
\end{table*}

The claim supported by the present construction is consequently
finite-domain and methodological. The reconstructed geometry, the
derivative-consistent reconstructed $f(R)$ sector, the reconstructed
source variables,
the radial null-energy diagnostic, the tidal-curvature diagnostic,
and the archived closure measure can be evaluated together on one
frozen numerical object. What is not established is a unique
microphysical matter model, a unique curvature--matter coupling,
or a complete covariant solution of a uniquely specified
nonminimally coupled $f(R)$ theory.

This hierarchy is used throughout the remainder of the manuscript.
Statements concerning geometry refer only to the Morris--Thorne
metric reconstruction; statements concerning $f_R$ and $f_{RR}$
refer only to the sampled curvature sector; statements concerning
$\rho$, $p_r$, and $p_t$ refer to reconstructed source variables;
and statements concerning the residual refer to the archived
source-side closure diagnostic defined in
Eq.~\eqref{eq:closure_diagnostic}. Stronger questions involving
exterior matching, global continuation, dynamical stability,
microphysical matter modelling, or a fully specified nonminimal
coupling remain outside the present finite-domain reconstruction.

\section{Analytically constrained Morris--Thorne geometry as the prescribed input}
\label{sec:wormhole_geometry}

The inverse reconstruction requires a geometrically admissible
prescribed input before the curvature sector is reconstructed.
Accordingly, the static, spherically symmetric Morris--Thorne ansatz
\cite{MorrisThorne1988,Visser1995} is parameterized so that the throat
condition, local flare-out condition, exterior radial no-horizon
condition, and boundedness of the redshift sector follow directly from
the functional form. These constraints therefore define the admissible
geometric input class prior to the curvature reconstruction. The dense-grid
quantities reported below are numerical cross-checks of the selected
geometry rather than acceptance conditions imposed after optimization.

\subsection{Metric ansatz, throat-radius normalization, and admissible radial domain}

The static, spherically symmetric line element is written as
\begin{equation}
ds^2
=
-e^{2\Phi(r)}dt^2
+
\frac{dr^2}{1-b(r)/r}
+
r^2d\Omega^2,
\label{eq:mt_metric}
\end{equation}
where
\begin{equation}
d\Omega^2
=
d\theta^2+\sin^2\theta\,d\phi^2.
\end{equation}
Here $\Phi(r)$ is the redshift function and $b(r)$ is the shape
function. The throat radius $r_0$ is defined by
\begin{equation}
b(r_0)=r_0.
\label{eq:throat_condition}
\end{equation}

The geometric length scale is normalized by the throat radius. If
$r_{\rm phys}$ and $b_{\rm phys}$ denote dimensionful radial and
shape-function quantities, the corresponding dimensionless
variables are
\begin{equation}
x
=
\frac{r_{\rm phys}}{r_{0,\rm phys}},
\qquad
\bar b
=
\frac{b_{\rm phys}}{r_{0,\rm phys}}.
\label{eq:throat_normalization}
\end{equation}
The numerical archive adopts throat-radius units and sets
\begin{equation}
r_0=1.
\end{equation}
For notational continuity with the numerical outputs, the
dimensionless radial coordinate is denoted by $r$ in the remainder
of the manuscript. Thus, the reported interval
$r\in[1,10^5]$ means $r_{\rm phys}/r_{0,\rm phys}\in[1,10^5]$;
it does not assign a physical value to the throat radius. A physical
length scale could later be restored through
\begin{equation}
r_{\rm phys}
=
r_{0,\rm phys}\,r,
\qquad
b_{\rm phys}
=
r_{0,\rm phys}\,b.
\label{eq:length_restoration}
\end{equation}
The ratios $b/r$, the redshift function $\Phi$, and the
no-horizon margin introduced below are dimensionless. Normalization
of curvature- and reconstructed-source-sector quantities is treated
separately where those quantities are introduced.

It is convenient to define
\begin{equation}
B(x)
\equiv
\frac{b(r)}{r},
\qquad
H(r)
\equiv
1-\frac{b(r)}{r}
=
1-B(x).
\label{eq:B_H_definition}
\end{equation}
The throat satisfies
\begin{equation}
H(r_0)=0,
\end{equation}
whereas a regular exterior radial metric requires
\begin{equation}
H(r)>0,
\qquad
r>r_0.
\label{eq:H_positive_condition}
\end{equation}
The local Morris--Thorne flare-out condition is
\begin{equation}
b'(r_0)<1.
\label{eq:flare_condition}
\end{equation}

\subsection{Shape-sector parameterization and analytic geometry conditions}

Rather than parameterizing the shape function directly, the
reconstruction uses the compactness profile
\begin{equation}
B(x)
=
x^{-\eta}\exp[-Q_b(x)],
\label{eq:B_ansatz}
\end{equation}
with
\begin{equation}
\begin{aligned}
Q_b(x)
&=
\left(\frac{x-1}{L_{\rm tail}}\right)^2
\operatorname{softplus}[N_b(x)],
\\
\operatorname{softplus}(z)
&=
\ln(1+e^z).
\end{aligned}
\label{eq:Qb_ansatz}
\end{equation}
and
\begin{equation}
b(r)=rB(x).
\end{equation}
Here $N_b(x)$ is the frozen shape-network output,
$\eta>0$ controls the local flare-out margin, and
$L_{\rm tail}>0$ sets the radial scale over which the compactness
profile departs from its throat value.

At $x=1$,
\begin{equation}
Q_b(1)=0,
\qquad
B(1)=1,
\end{equation}
and hence
\begin{equation}
b(r_0)=r_0
\end{equation}
holds analytically. Provided that $N_b(x)$ is finite and
differentiable at the throat, the quadratic prefactor in
Eq.~\eqref{eq:Qb_ansatz} also gives
\begin{equation}
Q_b'(1)=0.
\end{equation}
Differentiating $b=rB(x)$ therefore yields
\begin{equation}
b'(r_0)
=
B(1)+B_x(1)
=
1-\eta.
\label{eq:analytic_bprime}
\end{equation}
Consequently,
\begin{equation}
1-b'(r_0)=\eta>0,
\end{equation}
so the local flare-out condition is a property of the
parameterization rather than a finite-difference acceptance test.

The same result can be expressed through the standard embedding
criterion \cite{MorrisThorne1988,Visser1995},
\begin{equation}
\left.
\frac{d^2r}{dz^2}
\right|_{r_0}
=
\left.
\frac{b-rb'}{2b^2}
\right|_{r_0}
=
\frac{\eta}{2r_0}
>0.
\label{eq:embedding_flare}
\end{equation}

The compactness parameterization also gives an analytic exterior
no-horizon result. Since
\begin{equation}
Q_b(x)\ge0
\end{equation}
and $\eta>0$, for every $x>1$,
\begin{equation}
0
<
B(x)
=
x^{-\eta}e^{-Q_b(x)}
\le
x^{-\eta}
<
1.
\label{eq:B_analytic_bound}
\end{equation}
It follows that
\begin{equation}
H(r)
=
1-B(x)
\ge
1-x^{-\eta}
>
0,
\qquad
r>r_0.
\label{eq:H_analytic_bound}
\end{equation}
Thus, within the parameterized geometry, the exterior radial
no-horizon condition follows analytically. The very small values of
$H$ immediately outside the throat in the floating-point export are
therefore numerical cross-checks of Eq.~\eqref{eq:H_analytic_bound},
rather than the basis of the no-horizon statement.

The same local expansion clarifies the radial coordinate behaviour
at the throat. From Eq.~\eqref{eq:analytic_bprime},
\begin{equation}
H(r)
=
\eta\frac{r-r_0}{r_0}
+
\mathcal O\!\left[(r-r_0)^2\right].
\label{eq:H_near_throat}
\end{equation}
The proper radial distance is
\begin{equation}
\ell(r)
=
\pm
\int_{r_0}^{r}
\frac{dr'}{\sqrt{H(r')}}.
\label{eq:proper_distance}
\end{equation}
Using Eq.~\eqref{eq:H_near_throat} gives
\begin{equation}
\ell(r)
=
\pm
2
\sqrt{
\frac{r_0(r-r_0)}{\eta}
}
+
\mathcal O\!\left[(r-r_0)^{3/2}\right],
\label{eq:proper_distance_throat}
\end{equation}
which is finite as $r\rightarrow r_0^+$. Therefore, the divergence
of the curvature-coordinate component $g_{rr}=1/H$ at the throat
does not introduce an additional infinite proper-distance
singularity.

Finally,
\begin{equation}
0<B(x)\le x^{-\eta}
\longrightarrow0
\qquad
(x\rightarrow\infty).
\label{eq:B_formal_tail}
\end{equation}
This limiting behaviour characterizes the metric parameterization
itself. It does not extend the numerical validation of the
reconstructed curvature and source sectors beyond their audited
finite domains.

From the inverse-problem perspective, these analytic relations remove
geometrically inadmissible candidates before the curvature-sector
reconstruction is evaluated, rather than penalizing throat or
no-horizon violations only through result-level numerical losses.
\subsection{Bounded redshift sector}

The selected geometry does not impose the ultrastatic condition
$\Phi=0$. Instead, the redshift profile is written as
\begin{equation}
\Phi(r)
=
A_\Phi
\tanh[N_\Phi(x)]
\exp[-Q_\Phi(x)],
\label{eq:Phi_ansatz}
\end{equation}
where
\begin{equation}
Q_\Phi(x)
=
\left(
\frac{x-x_c}{w_\Phi}
\right)^2,
\qquad
w_\Phi>0.
\label{eq:QPhi_ansatz}
\end{equation}
Here $N_\Phi(x)$ is the frozen redshift-network output,
$A_\Phi$ is a finite amplitude, $w_\Phi$ is a positive width, and
$x_c$ sets the centre of the localized redshift structure.

Because
\begin{equation}
|\tanh[N_\Phi(x)]|\le1,
\qquad
0<e^{-Q_\Phi(x)}\le1,
\end{equation}
the redshift function satisfies the analytic bound
\begin{equation}
|\Phi(r)|
\le
|A_\Phi|
<
\infty.
\label{eq:Phi_bound}
\end{equation}
Accordingly,
\begin{equation}
e^{-2|A_\Phi|}
\le
e^{2\Phi(r)}
\le
e^{2|A_\Phi|},
\label{eq:temporal_metric_bound}
\end{equation}
so the temporal metric coefficient remains finite and strictly
nonzero wherever the parameterization is evaluated.

For finite $x_c$ and $w_\Phi>0$,
\begin{equation}
Q_\Phi(x)\rightarrow\infty,
\qquad
\Phi(x)\rightarrow0
\qquad
(x\rightarrow\infty).
\label{eq:Phi_formal_tail}
\end{equation}
Together with Eq.~\eqref{eq:B_formal_tail}, this means that the
chosen metric functions possess a formal asymptotically flat tail at
the level of the geometric ansatz. This statement concerns the metric
parameterization only; it does not establish a globally continued
solution of the reconstructed modified-gravity and source sectors.

\subsection{Frozen-grid verification of the prescribed geometry}

After the analytic geometric constraints are fixed, the prescribed
geometry used by the inverse reconstruction is evaluated on the common
finite radial grid
\begin{equation}
\mathcal D_r
=
\{r_i\}_{i=1}^{10019},
\qquad
r_i\in[1,10^5],
\qquad
r_0=1.
\label{eq:geometry_grid}
\end{equation}
On this grid the throat node satisfies
\begin{equation}
b(r_0)-r_0=0.
\end{equation}
The exported exterior no-horizon diagnostic has the minimum positive
floating-point value
\begin{equation}
\min_{r_i>r_0}^{+}H(r_i)
=
2.220446\times10^{-16}.
\label{eq:H_numeric_crosscheck}
\end{equation}
Because Eq.~\eqref{eq:H_analytic_bound} already establishes
$H(r)>0$ for $r>r_0$ within the parameterized geometry, the value in
Eq.~\eqref{eq:H_numeric_crosscheck} is interpreted only as a
finite-precision cross-check near the throat.

The exported redshift profile satisfies
\begin{equation}
\begin{aligned}
\Phi_{\min}
&=
-3.139897645936\times10^{-4},
\\
\Phi_{\max}
&=
9.920371941187\times10^{-2}.
\end{aligned}
\label{eq:Phi_numeric_range}
\end{equation}
with
\begin{equation}
\begin{aligned}
\max_i|\Phi(r_i)|
&=
9.920371941187\times10^{-2},
\\
\min_i e^{2\Phi(r_i)}
&=
9.993722176087\times10^{-1}.
\end{aligned}
\label{eq:Phi_numeric_crosscheck}
\end{equation}
These values are consistent with the analytic boundedness of
Eqs.~\eqref{eq:Phi_bound}--\eqref{eq:temporal_metric_bound}.
\begin{figure}[tbp]
\centering
\includegraphics[width=\columnwidth]{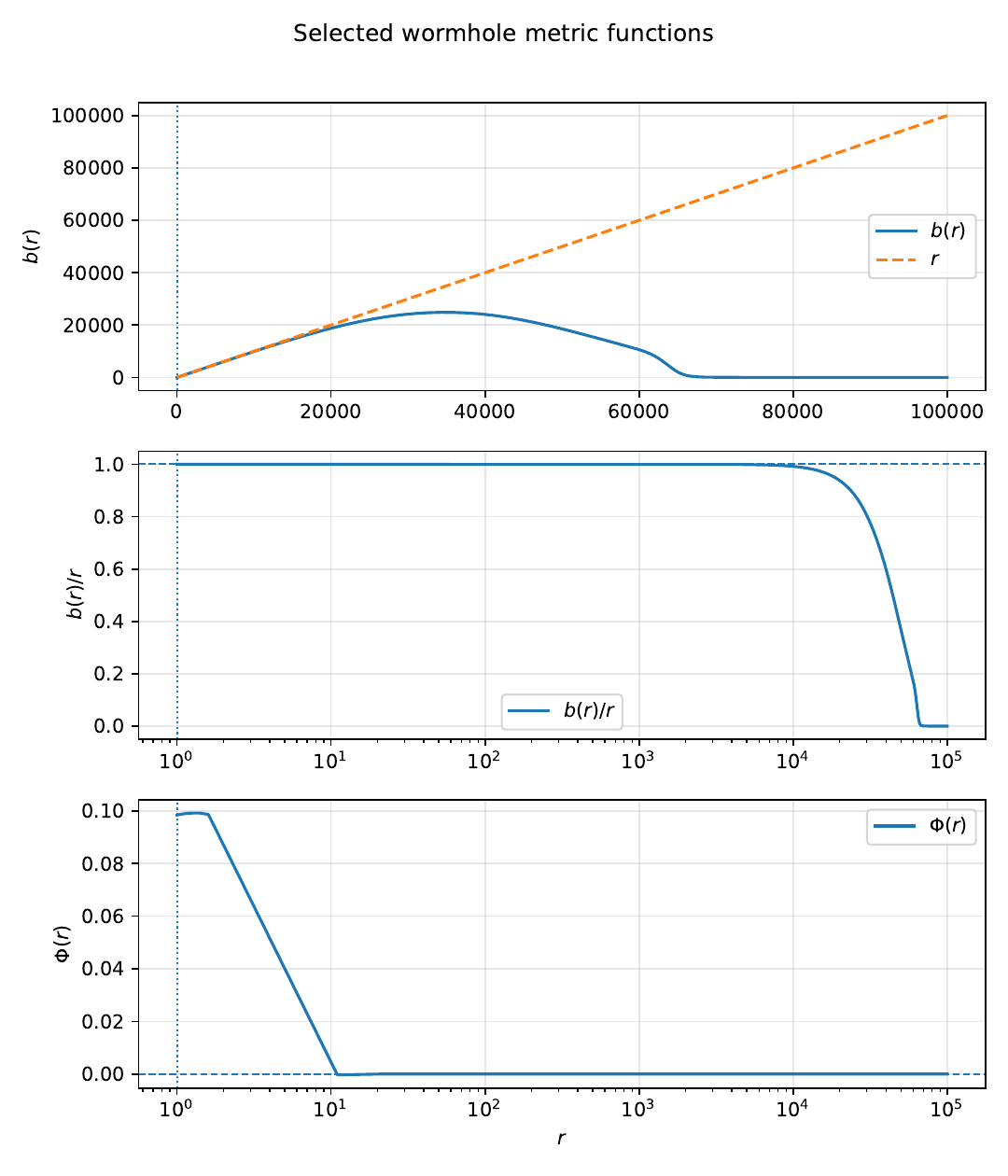}
\caption{
Metric profiles of the prescribed Morris--Thorne-type geometry used as
the geometric input to the inverse reconstruction.
The panels show the shape function $b(r)$, the compactness
$b(r)/r$, and the bounded redshift function $\Phi(r)$ on the
finite numerical domain. Radial coordinates are reported in
throat-radius units, with $r_0=1$. The throat and local flare-out
conditions follow analytically from the compactness
parameterization, while the exterior condition $H(r)>0$ follows
from Eq.~\eqref{eq:H_analytic_bound}. The plotted profiles provide
finite-grid cross-checks of these analytic geometric properties.
}
\label{fig:metric_functions}
\end{figure}

Figure~\ref{fig:metric_functions} displays the selected metric
profiles, while the detailed throat, flare-out, radial no-horizon,
redshift, and grid diagnostics are retained in the Supplementary
Material. The geometric statement established here is therefore
specific and limited: the adopted metric parameterization contains
an exact throat, analytic local flare-out, analytic exterior
$H(r)>0$, bounded nonvanishing temporal metric coefficient, and a
locally finite proper radial distance at the throat. These geometric
properties do not by themselves establish exterior matching,
dynamical stability, or observer-dependent safe traversal.


\section{Derivative-consistent inverse reconstruction of the $f(R)$ sector}
\label{sec:fr_reconstruction}

The prescribed geometry of Sec.~3 determines the sampled
Ricci-scalar trajectory on which the curvature-sector inverse
reconstruction is performed. The unknown curvature hierarchy is not
represented by three independently adjustable dense-grid arrays.
Instead, the reconstruction is restricted to an admissible functional
class generated from a single analytic representation of
$f_{RR}(R)$, with $f_R(R)$ and $f(R)$ obtained by successive
integration. In this way, derivative consistency is imposed at the
level of the reconstruction map itself rather than assessed only after
numerical fitting. The sampled trajectory is used directly, so the
method does not require construction of a global inverse $r=r(R)$.
The resulting curvature object is then subjected to the finite-domain
sign and conditioning diagnostics defined below.

\subsection{Sampled curvature trajectory and reconstruction domain}
\label{subsec:curvature_trajectory_domain}

Let
\begin{equation}
\mathcal T_R
=
\left\{
(r_i,R_i):
R_i=R(r_i),
\quad
r_i\in\mathcal D_r
\right\},
\label{eq:curvature_trajectory_def}
\end{equation}
with
\begin{equation}
\mathcal D_r
=
\{r_i\}_{i=1}^{10019},
\qquad
r_i\in[1,10^5].
\label{eq:curvature_radial_grid}
\end{equation}
The sampled curvature interval is
\begin{equation}
\begin{aligned}
\mathcal I_R&=[R_{\min},R_{\max}],
\\
R_{\min}&=-1.1317148307\times10^{-9},
\\
R_{\max}&=1.9999999979 .
\end{aligned}
\label{eq:curvature_sampled_interval}
\end{equation}

The interval $\mathcal I_R$ is the image of the audited finite
radial grid under the selected geometry. It is therefore the
curvature interval on which the numerical reconstruction is
assessed; it is not asserted to be a global curvature domain for
an unrestricted $f(R)$ theory.

No global monotonicity assumption for $R(r)$ is required by the
reconstruction. In particular, the numerical procedure does not
construct or use an inverse map $r=r(R)$. The curvature functions
are defined directly as single-valued functions of $R$. Hence, if
two radial locations sample the same curvature,
\begin{equation}
R(r_a)=R(r_b),
\label{eq:repeated_R_condition}
\end{equation}
the formula-first reconstruction necessarily assigns
\begin{equation}
\begin{aligned}
f(R(r_a))      &= f(R(r_b)),\\
f_R(R(r_a))    &= f_R(R(r_b)),\\
f_{RR}(R(r_a)) &= f_{RR}(R(r_b)).
\end{aligned}
\label{eq:repeated_R_formula_consistency}
\end{equation}
Thus, possible repeated curvature values along the radial trajectory
do not generate multiple curvature-sector branches through a
numerical inversion.

The radial coordinate has already been normalized by the physical
throat radius in Sec.~\ref{sec:wormhole_geometry}. For dimensional
interpretation of the curvature sector, we adopt the corresponding
throat-scaled reporting convention
\begin{equation}
\bar R
=
r_{0,\mathrm{phys}}^{\,2}R_{\mathrm{phys}},
\qquad
\bar f
=
r_{0,\mathrm{phys}}^{\,2}f_{\mathrm{phys}},
\label{eq:curvature_scaling_convention}
\end{equation}
so that
\begin{equation}
\bar f_R
=
\frac{d\bar f}{d\bar R}
=
f_{R,\mathrm{phys}},
\qquad
\bar f_{RR}
=
\frac{d^2\bar f}{d\bar R^2}
=
\frac{f_{RR,\mathrm{phys}}}
     {r_{0,\mathrm{phys}}^{\,2}}.
\label{eq:curvature_derivative_scaling}
\end{equation}
For notational economy, the bars are suppressed below. Accordingly,
the numerical curvature-sector quantities reported in this work
refer to the throat-scaled reconstruction rather than to an
independently assigned SI curvature scale.

\subsection{Positive analytic generator and derivative-consistent integration hierarchy}
\label{subsec:canonical_fr_generator}

The reconstructed second derivative is represented by the positive
analytic form
\begin{equation}
f_{RR}(R)
=
f_{\mathrm{floor}}
+
\sum_{j=1}^{N_G}
w_j
\exp\!\left[
-\frac{1}{2}
\left(
\frac{R-c_j}{s_j}
\right)^2
\right],
\label{eq:canonical_fRR_generator}
\end{equation}
with
\begin{equation}
f_{\mathrm{floor}}>0,
\qquad
w_j\ge0,
\qquad
s_j>0.
\label{eq:canonical_fRR_constraints}
\end{equation}
Consequently,
\begin{equation}
f_{RR}(R)>0
\qquad
(R\in\mathcal I_R)
\label{eq:canonical_fRR_analytic_positive}
\end{equation}
for the selected analytic branch.

For an archived curvature-space integration anchor $R_\star$, the
lower-order functions are defined by
\begin{equation}
f_R(R)
=
f_R(R_\star)
+
\int_{R_\star}^{R}
f_{RR}(u)\,du ,
\label{eq:canonical_fR_integral}
\end{equation}
and
\begin{equation}
f(R)
=
f(R_\star)
+
\int_{R_\star}^{R}
f_R(u)\,du .
\label{eq:canonical_f_integral}
\end{equation}
Therefore,
\begin{equation}
\frac{df}{dR}=f_R,
\qquad
\frac{df_R}{dR}=f_{RR},
\label{eq:canonical_curvature_identities}
\end{equation}
identically by construction.

This construction reduces the derivative hierarchy to one generated
functional degree of freedom together with the integration anchors
required to recover the lower-order members. It therefore excludes,
by construction, candidates in which nominally reconstructed
$f$, $f_R$, and $f_{RR}$ profiles are mutually incompatible as
derivatives of a common function. The restriction is an admissibility
constraint on the reconstructed function class; it is not a uniqueness
statement for the inverse problem.

The final frozen curvature arrays used in the validation layer are
obtained only by evaluating this same analytic reconstruction at the
sampled values $R_i$. They are not produced by independently fitting
or subsequently correcting $f$, $f_R$, or $f_{RR}$ after candidate
selection. The complete numerical parameter vector of the selected
generator and its integration anchors is retained in the frozen
machine-readable reconstruction archive, so that the same functional
object can be re-evaluated independently of the plotted dense-grid
arrays.

\subsection{Finite-domain admissibility and conditioning diagnostics}
\label{subsec:canonical_curvature_diagnostics}

The final derivative-consistent curvature object contains
$N=10019$ sampled radial nodes. On this object, the second
curvature derivative satisfies
\begin{equation}
\min_i f_{RR}(R_i)
=
0.0678574418696333
>0,
\label{eq:canonical_fRR_min}
\end{equation}
and
\begin{equation}
\max_i f_{RR}(R_i)
=
980.430287696675 .
\label{eq:canonical_fRR_max}
\end{equation}
The same final audit gives
\begin{equation}
\min_i f_R(R_i)>0,
\qquad
\min_i f_{RR}(R_i)>0.
\label{eq:canonical_fr_sign_gates}
\end{equation}

These conditions are used here only as local
curvature-sector viability diagnostics on the sampled trajectory,
as customary in $f(R)$ gravity
\cite{SotiriouFaraoni2010,DeFeliceTsujikawa2010}.
They do not establish perturbative stability, global stability,
uniqueness, or viability outside $\mathcal I_R$. In particular,
no statement about the sign of a candidate-specific nonminimal
prefactor is inferred from Eq.~\eqref{eq:canonical_fr_sign_gates},
because such a prefactor is not uniquely instantiated in the
present reconstruction.

\begin{figure*}[tbp]
\centering
\includegraphics[width=\textwidth]{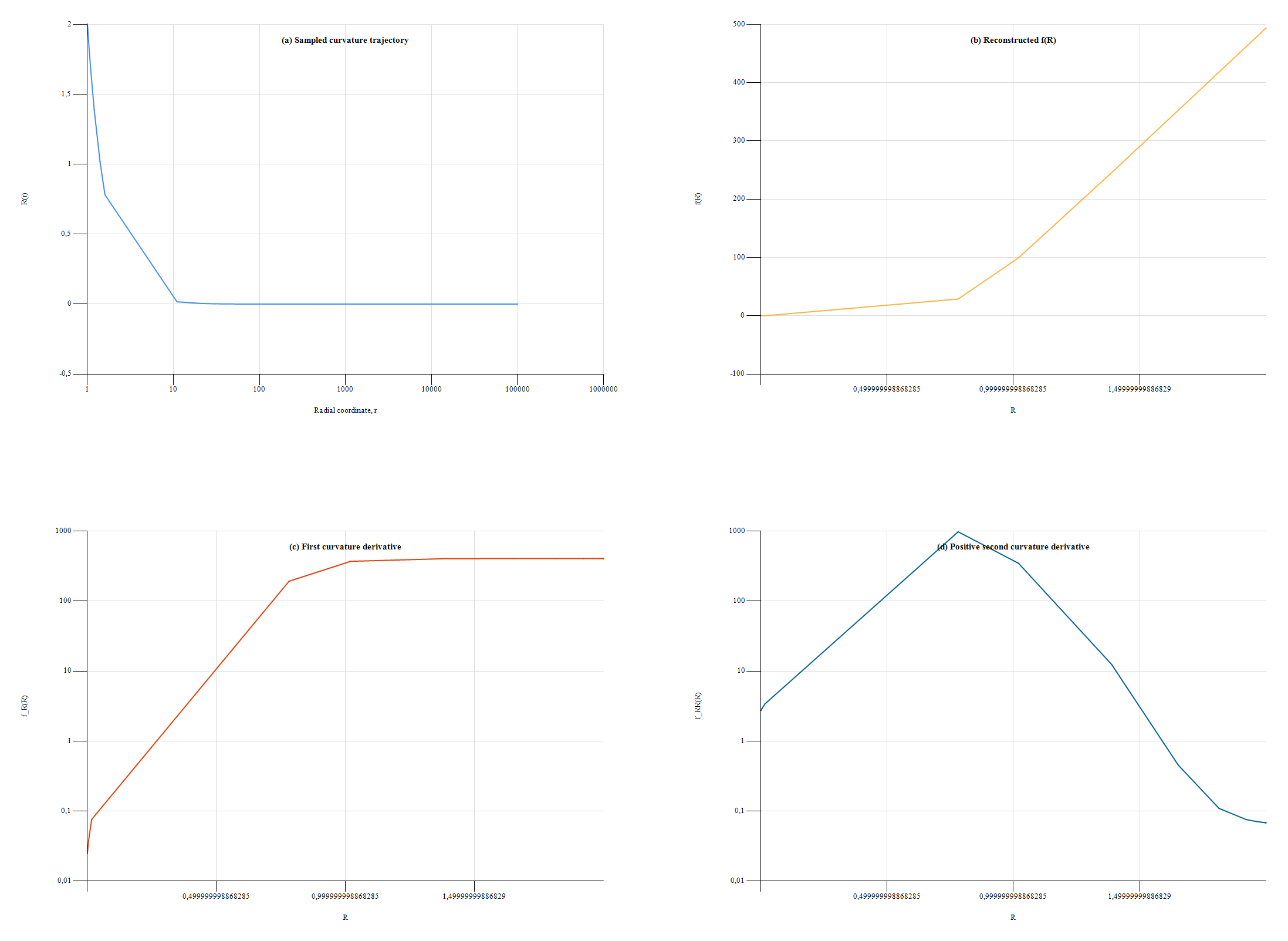}
\caption{
Sampled curvature trajectory and derivative-consistent inverse
reconstruction of the $f(R)$ sector for the final frozen finite-domain
object.
The panels show the sampled $R(r)$ trajectory together with
$f(R)$, $f_R(R)$, and $f_{RR}(R)$ evaluated from one analytic
curvature reconstruction. The $f_{RR}$ panel uses logarithmic
vertical scaling because the final frozen profile spans a broad
positive range. No inverse mapping $r=r(R)$ is used.
All curves correspond to the same frozen curvature arrays used in
the numerical validation.
}
\label{fig:fr_reconstruction}
\end{figure*}

Figure~\ref{fig:fr_reconstruction} displays the same final frozen curvature arrays used in
Eqs.~(69)--(71). The $f_{RR}$ panel is shown on a logarithmic ordinate so that the full
positive sampled dynamic range remains visible.

The curvature-sector conclusion is therefore finite-domain and
object-specific. Within the sampled reconstruction domain, the selected
functional object belongs to the imposed derivative-consistent class
and satisfies $f_R>0$ and $f_{RR}>0$ at every audited node. These
properties establish admissibility of the reported branch under the
criteria used in this work; they do not imply uniqueness of the inverse
reconstruction, a global $f(R)$ stability theorem, exterior completion,
or a uniquely specified nonminimal curvature--matter-coupling solution.

\section{Frozen-object validation and provenance protocol}
\label{sec:same_run_validation_protocol}

The geometry and derivative-consistent curvature sector are fixed
before any result-level diagnostic is evaluated. The validation stage
therefore acts on a post-selection frozen object rather than providing
additional degrees of freedom to the reconstruction. Its purpose is to
verify that the source-side closure, reconstructed-source quantities,
radial null-energy data, tidal profile, and arithmetic recomputation
are all associated with the same numerical provenance.

This separation between reconstruction and validation is important for
the inverse-problem interpretation. A small training loss or a
post-processed residual does not by itself establish that the reported
diagnostics belong to one structurally admissible reconstruction
\cite{RaissiEtAl2019,WangTengPerdikaris2020,CuomoEtAl2022}.
Accordingly, no metric, curvature, source, closure, null-energy, or
tidal quantity is refitted after the final reconstruction has been
selected. Historical archive files occasionally use the term
``same-run'' for this common-provenance requirement; here it denotes
only shared numerical provenance and not an independent covariant
re-solution of a uniquely specified curvature--matter-coupled theory.

\subsection{Frozen reconstruction object and no-post-selection rule}
\label{subsec:frozen_candidate_policy}

All validation quantities are tied to the common radial grid
$\mathcal D_r$ defined in Eq.~\eqref{eq:geometry_grid}. After candidate
selection, the complete numerical object supplied to the validation
layer is represented as
\begin{equation}
\begin{aligned}
\mathcal C_{\rm frz}
=
\{&
r_i,b_i,\Phi_i,H_i,
R_i,f_i,f_{R,i},f_{RR,i},
\\
&\rho_i,p_{r,i},p_{t,i},
\mathcal T_i,
Q^{\rm base}_{c,i},
Q^{\rm src}_{c,i}
\}_{i=1}^{N},
\end{aligned}
\label{eq:frozen_candidate_object}
\end{equation}
with
\begin{equation}
c\in\{tt,rr,\theta\theta,{\rm trace}\}.
\end{equation}
Here
\begin{equation}
H_i=1-\frac{b_i}{r_i},
\qquad
R_i=R(r_i),
\qquad
f_i=f(R_i),
\end{equation}
and $f_{R,i}$ and $f_{RR,i}$ are evaluated from the same
reconstructed curvature-sector object described in
Sec.~\ref{sec:fr_reconstruction}.

The quantities $Q^{\rm base}_{c,i}$ and $Q^{\rm src}_{c,i}$ are the two
archived channel arrays used in the source-side closure comparison
defined in Eq.~\eqref{eq:closure_diagnostic}. No candidate-specific
value of $\lambda$, $h(R)$, $\mathcal L_m$, or $F_{\rm NMC}(r)$ is
added to the frozen object at this stage.

Once $\mathcal C_{\rm frz}$ has been selected, none of the metric,
curvature, reconstructed-source, closure-channel, radial null-energy,
or tidal arrays is altered in response to a subsequent diagnostic.
The validation map is therefore evaluated on fixed inputs. In
particular, failure of a later diagnostic would reject or qualify the
selected reconstruction rather than trigger diagnostic-specific
refitting. This no-post-selection rule is the operational meaning of
``frozen'' throughout the validation protocol.

\subsection{Archived source-side closure as a validation functional}
\label{subsec:channelwise_field_equation_residuals}

The numerical closure audit uses the four archived channels
\begin{equation}
\mathcal C_{\rm ch}
=
\{tt,rr,\theta\theta,{\rm trace}\}.
\label{eq:residual_channel_set_5_2}
\end{equation}
For each channel, the closure difference is
\begin{equation}
\mathcal C_c(r_i)
=
Q^{\rm base}_c(r_i)
-
Q^{\rm src}_c(r_i),
\qquad
c\in\mathcal C_{\rm ch},
\label{eq:channel_residual_definition_5_2}
\end{equation}
consistent with Eq.~\eqref{eq:closure_diagnostic}.  The associated
finite-grid maximum norm is
\begin{equation}
\|\mathcal C\|_{\infty,\mathcal D_r}
=
\max_{c\in\mathcal C_{\rm ch}}
\max_{r_i\in\mathcal D_r}
|\mathcal C_c(r_i)|.
\label{eq:residual_norm_5_2}
\end{equation}

The absolute closure values have the same numerical normalization
as their corresponding archived baseline and source-side channel
arrays.  Because the present reconstruction does not assign an
independently restored SI scale to these channel quantities, the
absolute closure magnitude is not interpreted as a dimensionful
covariant residual in physical units.  A scale-independent measure
is instead obtained from
\begin{equation}
\epsilon_{\mathcal C}
=
\frac{
\|\mathcal C\|_{\infty,\mathcal D_r}
}{
\|Q^{\rm base}\|_{\infty,\mathcal D_r}
},
\qquad
\|Q^{\rm base}\|_{\infty,\mathcal D_r}
=
\max_{c,i}|Q^{\rm base}_c(r_i)|,
\label{eq:relative_closure_measure}
\end{equation}
with the equivalent percentage reduction
\begin{equation}
\eta_{\mathcal C}
=
100\left(1-\epsilon_{\mathcal C}\right)\%.
\label{eq:closure_reduction_measure}
\end{equation}
The quantities in Eqs.~\eqref{eq:residual_channel_set_5_2}--%
\eqref{eq:closure_reduction_measure} are evaluated only after the
reconstruction has been frozen. They therefore function as
post-selection validation measures rather than terms that are adjusted
to improve the reported reconstruction. Their numerical values are
reported in Sec.~\ref{sec:results}; the present subsection defines only their
construction and interpretation.

Importantly, this validation functional compares the two archived
source-side channel representations already contained in the frozen
object. It is not identified with an independent covariant evaluation
of the reference field equation in Eq.~\eqref{eq:reference_nmc_field_equation}
for a uniquely specified set $\{\lambda,h(R),\mathcal L_m\}$.
Consequently, a small value of the closure functional establishes
internal numerical agreement of the archived representation, not
completion of a microscopic gravitational model.

\begin{figure}[tbp]
\centering
\includegraphics[width=0.86\linewidth]{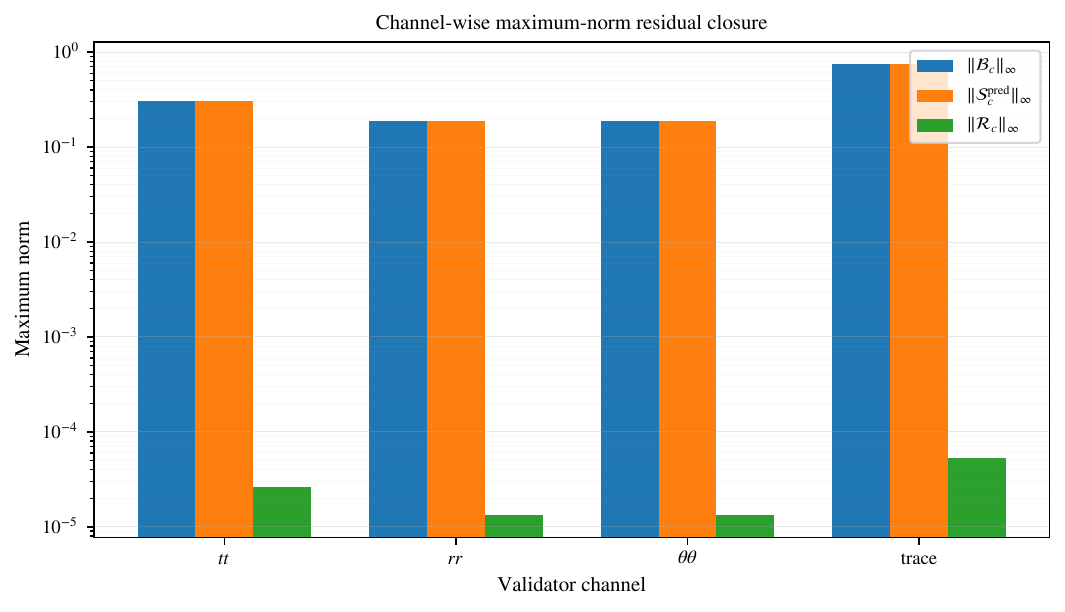}
\caption{
Post-selection maximum-norm comparison of the four archived closure
channels evaluated on the frozen reconstruction.  In the historical plotting notation,
$\mathcal B_c$, $\mathcal S^{\rm pred}_c$, and
$\mathcal R_c$ correspond respectively to the archived baseline
$Q^{\rm base}_c$, the archived source-side prediction
$Q^{\rm src}_c$, and their difference
$\mathcal C_c=Q^{\rm base}_c-Q^{\rm src}_c$.
The figure summarizes internal numerical closure of the archived
object; it is not an independent covariant solution test of a
uniquely specified nonminimal theory.
}
\label{fig:residual_absmax_reduction_5_2}
\end{figure}

Figure~\ref{fig:residual_absmax_reduction_5_2} is retained because
it provides a compact comparison of the archived channel scales and
their closure differences without introducing an additional fitted
quantity.

\FloatBarrier

\subsection{Reconstructed source variables and interpretation}
\label{subsec:source_sector_export_physical_interpretation}

The same frozen validation object also contains the three
reconstructed source variables $\rho_i$, $p_{r,i}$, and $p_{t,i}$.  For the finite-domain
diagnostics they are collected in the orthonormal-frame
representation
\begin{equation}
T^{({\rm rec})}_{\hat\mu\hat\nu}(r_i)
=
\operatorname{diag}
(\rho_i,p_{r,i},p_{t,i},p_{t,i}),
\qquad
r_i\in\mathcal D_r .
\label{eq:source_export_orthonormal_tensor_5_3}
\end{equation}

The superscript ``rec'' emphasizes their status in this work:
these quantities are reconstructed source variables, not a
stress--energy tensor obtained by varying a unique microscopic
matter Lagrangian.  Accordingly, the later positivity tests applied
to combinations of $\rho$, $p_r$, and $p_t$ characterize this
reconstructed source representation only.

This distinction is essential for the interpretation of exoticity.
In explicitly specified curvature--matter-coupled theories, the
division of wormhole support between matter, curvature, and coupling
terms depends on the action and matter Lagrangian
\cite{BertolamiEtAl2008,HarkoLobo2010,
GarciaLobo2010a,GarciaLobo2010b}.  Because those microscopic
ingredients are not uniquely reconstructed here, the archived
source-side closure cannot be used to conclude that all possible
exotic support has been eliminated or uniquely assigned to one
physical sector.

The numerical values of $\rho$, $p_r$, and $p_t$ are quoted in the
native source normalization of the frozen archive.  No conversion
to SI energy-density or pressure units is performed in the present
finite-domain reconstruction.  A dimensional interpretation would
require, in addition to the throat-radius scale, a fully specified
gravitational/source normalization and matter model.  The
pointwise combinations and coordinate-radial null-energy diagnostic
constructed from these variables are defined and evaluated in
Sec.~\ref{sec:physical_viability_diagnostics} rather than repeated
here.

\subsection{Internal recomputation and provenance verification}
\label{subsec:reproducibility_recomputation_checks}

The archive-level recomputation check asks whether the stored
closure columns can be regenerated directly from the same frozen
baseline and source-side arrays.  Define
\begin{equation}
\widehat{\mathcal C}_c(r_i)
=
Q^{\rm base}_c(r_i)
-
Q^{\rm src}_c(r_i),
\label{eq:recomputed_closure_5_4}
\end{equation}
and
\begin{equation}
\Delta_{\rm rec}^{\max}
=
\max_{i,c}
\left|
\widehat{\mathcal C}_c(r_i)
-
\mathcal C^{\rm arch}_c(r_i)
\right|.
\label{eq:recompute_mismatch_channel_5_4}
\end{equation}
This test detects inconsistencies between the archived input channels
and the stored closure column.  It is therefore an arithmetic and
provenance check, not an independent physical solution test.

The frozen archive additionally contains the radial grid, metric
profiles, curvature trajectory, reconstructed curvature arrays,
reconstructed source variables, radial null-energy data, tidal
profile, closure-channel arrays, machine-readable summaries,
figure-generation inputs, and file checksums.  These materials
support deterministic internal recomputation of the reported
diagnostics.

Internal recomputation must be distinguished from independent
third-party reproduction.  A roundoff-level value of
$\Delta_{\rm rec}^{\max}$ demonstrates consistency of the archived
columns but does not establish that an external researcher has
reproduced the complete reconstruction from independently obtained
inputs.  Public, versioned release of the frozen arrays, scripts,
environment information, and checksums is therefore treated as a
separate reproducibility requirement rather than as a consequence
of the internal recomputation test.

Thus, the validation hierarchy distinguishes three logically different
claims: structural admissibility of the reconstructed functional
object, internal consistency of diagnostics evaluated on that frozen
object, and independent external reproduction. Only the first two are
demonstrated numerically in the present study; the public archive makes
the third test possible but does not claim that it has already been
performed.


\section{Post-selection finite-domain source and tidal diagnostics}
\label{sec:physical_viability_diagnostics}

All diagnostics in this section are evaluated after the numerical
reconstruction has been frozen according to the protocol of
Sec.~\ref{sec:same_run_validation_protocol}. They therefore probe
properties of one fixed reconstruction rather than supplying
additional degrees of freedom to the inverse problem. In particular,
$\rho$, $p_r$, and $p_t$ are the reconstructed source variables
introduced in
Sec.~\ref{subsec:source_sector_export_physical_interpretation};
they are not interpreted as a stress--energy tensor obtained by
varying a uniquely specified microscopic matter Lagrangian.

The post-selection diagnostic layer considered here contains three
logically distinct components: pointwise algebraic energy-condition
combinations of the reconstructed source variables, a cumulative
finite-domain coordinate-radial null-energy integral, and an archived
validator-normalized tidal-curvature profile for the prescribed
geometry. These quantities test different aspects of the frozen
object and are not interchangeable. In particular, positivity of the
reconstructed-source combinations does not determine the tidal
curvature, and neither diagnostic identifies a unique microscopic
division of wormhole support among matter, curvature, and
nonminimal-coupling terms.

The reconstructed source variables retain the native numerical
normalization of the frozen archive, as discussed in
Sec.~\ref{subsec:source_sector_export_physical_interpretation}.
Their signs and algebraic combinations can therefore be audited
directly on the finite grid, but no independent SI energy-density or
pressure scale is inferred. Likewise, the tidal quantity used below
retains its archived validator normalization. The interpretation of
each diagnostic is therefore restricted to the numerical
normalization in which the frozen reconstruction was validated.

\subsection{Pointwise reconstructed-source energy-condition audit}
\label{subsec:pointwise_energy_conditions}

For compact notation, the reconstructed source variables are
collected in the orthonormal-frame representation
\begin{equation}
T^{({\rm rec})}_{\hat\mu\hat\nu}(r_i)
=
\operatorname{diag}
\left(
\rho_i,p_{r,i},p_{t,i},p_{t,i}
\right),
\qquad
r_i\in\mathcal D_r .
\label{eq:ec_orthonormal_source_6_1}
\end{equation}
The superscript ``rec'' is essential: the object in
Eq.~\eqref{eq:ec_orthonormal_source_6_1} is a reconstructed
source representation, not a microphysically derived matter tensor.

For a diagonal anisotropic source representation, the standard
pointwise energy-condition combinations are
\cite{HawkingEllis1973,Wald1984,KontouSanders2020}
\begin{equation}
\begin{aligned}
M_{\rho,i}
&=\rho_i,
\\
M_{{\rm NEC},r,i}
&=\rho_i+p_{r,i},
\\
M_{{\rm NEC},t,i}
&=\rho_i+p_{t,i},
\\
M_{{\rm SEC},i}
&=\rho_i+p_{r,i}+2p_{t,i},
\\
M_{{\rm DEC},r,i}
&=\rho_i-|p_{r,i}|,
\\
M_{{\rm DEC},t,i}
&=\rho_i-|p_{t,i}|.
\end{aligned}
\label{eq:ec_margins_6_1}
\end{equation}

On the common $N=10019$ node grid, the minimum values are
\begin{equation}
\begin{aligned}
\min_i M_{\rho,i}
&=6.000000\times10^{-2},
\\
\min_i M_{{\rm NEC},r,i}
&=8.000000\times10^{-2},
\\
\min_i M_{{\rm NEC},t,i}
&=8.000000\times10^{-2},
\\
\min_i M_{{\rm SEC},i}
&=1.195736\times10^{-1},
\\
\min_i M_{{\rm DEC},r,i}
&=4.000000\times10^{-2},
\\
\min_i M_{{\rm DEC},t,i}
&=4.000000\times10^{-2}.
\end{aligned}
\label{eq:ec_minimum_values_6_1}
\end{equation}
These minima are evaluated directly from the reconstructed-source
arrays contained in $\mathcal C_{\rm frz}$; no source variable is
adjusted after freezing in order to enforce their positivity.

Thus, every listed reconstructed-source combination remains
strictly positive on the audited finite grid. In the usual
energy-condition terminology, the reconstructed triplet satisfies
the corresponding WEC, radial NEC, tangential NEC, SEC, radial DEC,
and tangential DEC inequalities on $\mathcal D_r$.

This statement is intentionally narrower than a claim that a
microphysical ``ordinary-matter'' model has been identified.
Because the present reconstruction does not determine a unique
matter Lagrangian or a unique curvature--matter coupling, positivity
of Eq.~\eqref{eq:ec_margins_6_1} cannot establish that all possible
exotic support has been eliminated from a fully specified theory.
It establishes only the algebraic properties of the archived
reconstructed source variables.

\begin{figure}[tbp]
\centering
\includegraphics[width=\linewidth]{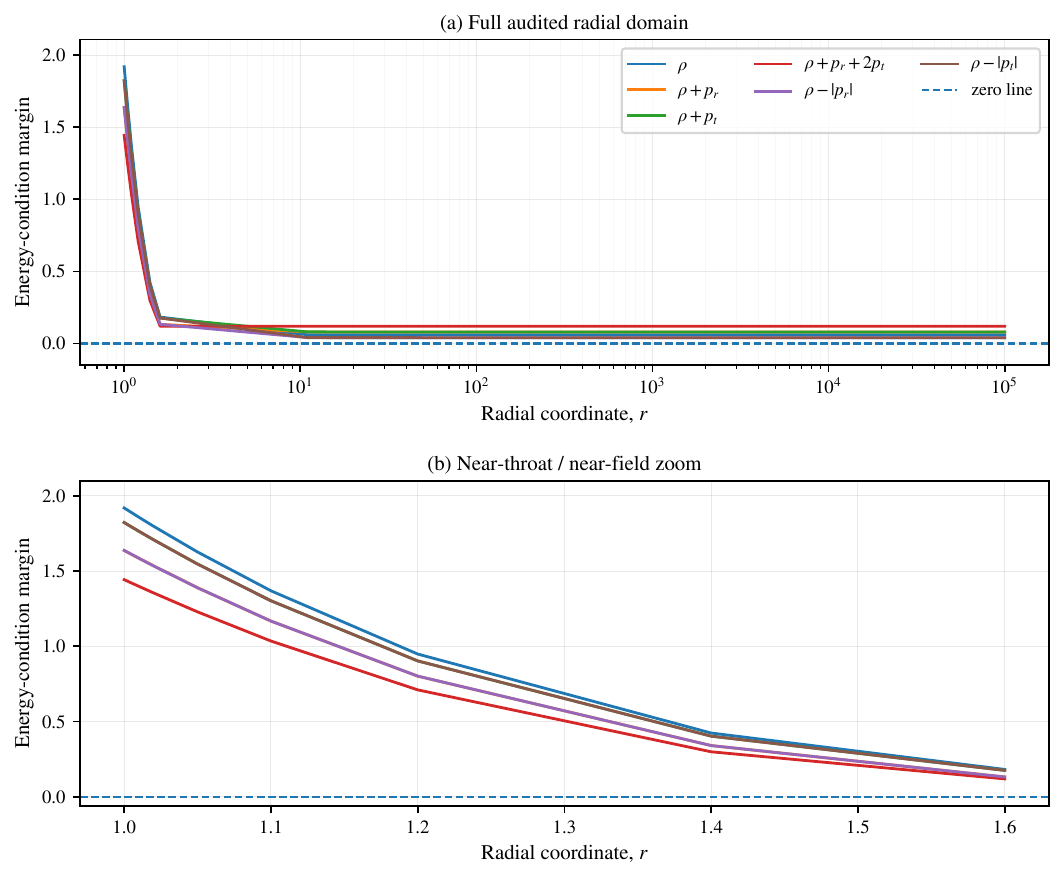}
\caption{
Post-selection pointwise energy-condition combinations constructed
from the reconstructed source variables on the frozen finite radial
grid.
The panels show $M_\rho$, $M_{{\rm NEC},r}$,
$M_{{\rm NEC},t}$, $M_{\rm SEC}$,
$M_{{\rm DEC},r}$, and $M_{{\rm DEC},t}$.
The zero reference line shows that every reported combination
remains positive on the audited grid. These curves characterize the
reconstructed source representation; they do not establish a unique
microphysical matter model.
}
\label{fig:energy_condition_margins_6_1}
\end{figure}

\FloatBarrier

\subsection{Finite-domain coordinate-radial null-energy audit}
\label{subsec:global_throat_band_anec_checks}

The radial null-energy combination constructed from the same
reconstructed source variables is
\begin{equation}
\mathcal N_r(r_i)
=
\rho_i+p_{r,i}.
\label{eq:radial_nec_integrand_6_2}
\end{equation}
To preserve compatibility with the archived data products and
figure labels, we retain the notation $I_{\rm ANEC}$ for the
cumulative quantity
\begin{equation}
I_{\rm ANEC}(r)
=
\int_{r_0}^{r}
\mathcal N_r(x)\,dx .
\label{eq:anec_radial_integrand_and_integral_6_2}
\end{equation}
However, $I_{\rm ANEC}$ in this manuscript is an archive label for
a \emph{coordinate-radial finite-domain integral}. It is not the
standard affine-parameter averaged null-energy integral over a
complete null geodesic
\cite{HawkingEllis1973,Wald1984,KontouSanders2020}.

This distinction also fixes its normalization. The coordinate $r$
is reported in throat-radius units and $\mathcal N_r$ carries the
native source normalization of the frozen archive. Consequently,
the numerical values below are finite-domain reconstruction
diagnostics rather than dimensionful, affine-invariant ANEC
quantities.

On the frozen grid,
\begin{equation}
\begin{aligned}
\min_i\mathcal N_r(r_i)
&=
8.000000\times10^{-2},
\\
\max_i\mathcal N_r(r_i)
&=
1.636502.
\end{aligned}
\label{eq:anec_integrand_bounds_6_2}
\end{equation}
Since $\mathcal N_r(r_i)>0$ at every audited node, the cumulative
coordinate-radial integral is monotone non-decreasing. The archived
full-domain and near-throat values are
\begin{equation}
\begin{aligned}
I_{\rm ANEC}(r_{\max})
&=
8000.52020616
>0,
\\
I_{\rm ANEC}(1.4)
&=
0.3522719805583363
>0.
\end{aligned}
\label{eq:anec_global_and_throat_values_6_2}
\end{equation}
Both values are reproduced by direct coordinate-trapezoidal
recomputation from the same frozen $\rho+p_r$ array. They are
therefore deterministic finite-grid functionals of the archived
reconstructed-source data rather than quantities obtained from an
additional fit or an independently adjusted source model.

\begin{figure}[tbp]
\centering
\includegraphics[width=\linewidth]{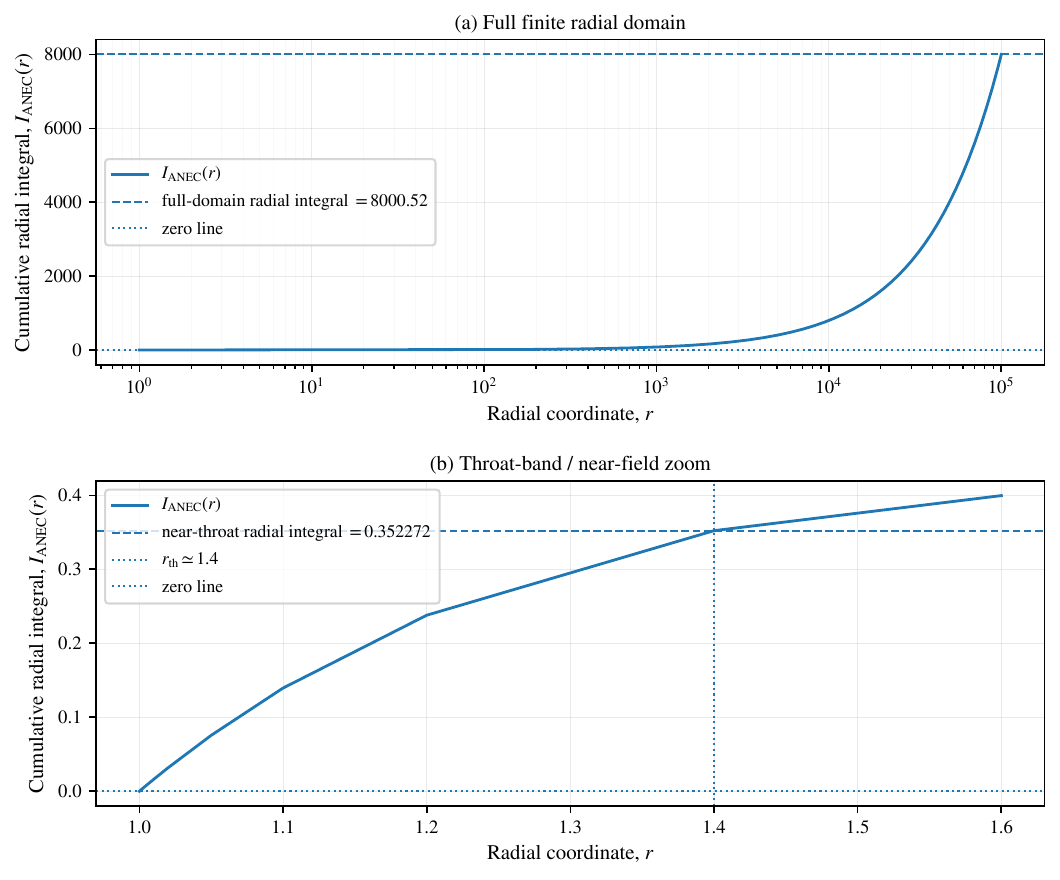}
\caption{
Post-selection cumulative finite-domain coordinate-radial
null-energy integral constructed from the reconstructed source
variables.
The upper panel shows
$I_{\rm ANEC}(r)=\int_{r_0}^{r}(\rho+p_r)\,dx$
over the complete audited interval, and the lower panel shows the
near-throat band $[r_0,1.4]$.
The symbol $I_{\rm ANEC}$ is retained as an archive label; the
plotted quantity is not an affine-complete averaged-null-energy
integral over a complete null geodesic.
}
\label{fig:cumulative_anec_6_2}
\end{figure}

\FloatBarrier

The conclusion supported by this calculation is therefore specific:
the reconstructed source representation has a positive radial NEC
combination at every audited node and positive cumulative
coordinate-radial integrals over both the full finite domain and the
selected near-throat band. No conclusion is drawn about
affine-complete ANEC, arbitrary null geodesics, exterior extensions,
or globally completed wormhole spacetimes.

\subsection{Archived validator-normalized tidal-curvature audit}
\label{subsec:tidal_force_traversability_audit}

The tidal-curvature audit probes the prescribed geometry as a
separate post-selection diagnostic and is logically independent of
the reconstructed-source inequalities. Positive reconstructed-source
energy-condition combinations do not, by themselves, constrain the
tidal curvature associated with the metric, and the tidal result is
therefore evaluated independently on the same frozen object.

For orientation, the static orthonormal radial and transverse
curvature components associated with
\begin{equation}
ds^2
=
-e^{2\Phi(r)}dt^2
+
\frac{dr^2}{H(r)}
+
r^2d\Omega^2,
\qquad
H(r)
=
1-\frac{b(r)}{r},
\label{eq:tidal_metric_form_6_3}
\end{equation}
may be written, up to the adopted Riemann-tensor sign convention, as
\begin{equation}
R_{\hat t\hat r\hat t\hat r}
=
H
\left[
\Phi''
+
(\Phi')^2
\right]
+
\frac{H'}{2}\Phi',
\label{eq:tidal_radial_component_6_3}
\end{equation}
and
\begin{equation}
R_{\hat t\hat\theta\hat t\hat\theta}
=
\frac{H\Phi'}{r},
\qquad
R_{\hat t\hat\phi\hat t\hat\phi}
=
R_{\hat t\hat\theta\hat t\hat\theta}.
\label{eq:tidal_transverse_component_6_3}
\end{equation}

These expressions provide the geometric context for interpreting
tidal curvature. The final numerical validation quantity, however,
is taken directly from the frozen archive as a dimensionless
validator-normalized profile,
\begin{equation}
\mathcal T_i
\equiv
\mathcal T(r_i),
\qquad
r_i\in\mathcal D_r .
\label{eq:archived_tidal_profile_6_3}
\end{equation}
The final archive identifies this profile itself as the numerical
source of the tidal diagnostic. The present validation therefore
does not reconstruct $\mathcal T_i$ from separately archived
orthonormal-component profiles and an independently specified
dimensional reference-curvature scale.

The normalization is defined so that
\begin{equation}
\mathcal T=1
\end{equation}
is the validator reference level. Accordingly,
\begin{equation}
\mathcal T<1
\label{eq:tidal_below_reference_6_3}
\end{equation}
means only that the archived normalized profile remains below that
reference level; it does not assign an observer-independent physical
safety threshold.

On the frozen grid, the archived profile gives
\begin{equation}
\begin{aligned}
\max_i\mathcal T(r_i)
&=
0.8665754053323194
<1,
\\
r_{\mathcal T,\max}
&\simeq
33203.988 .
\end{aligned}
\label{eq:tidal_max_value_location_6_3}
\end{equation}
The maximum occurs in the extended radial region rather than in the
immediate throat neighbourhood.

\begin{figure}[tbp]
\centering
\includegraphics[width=\linewidth]{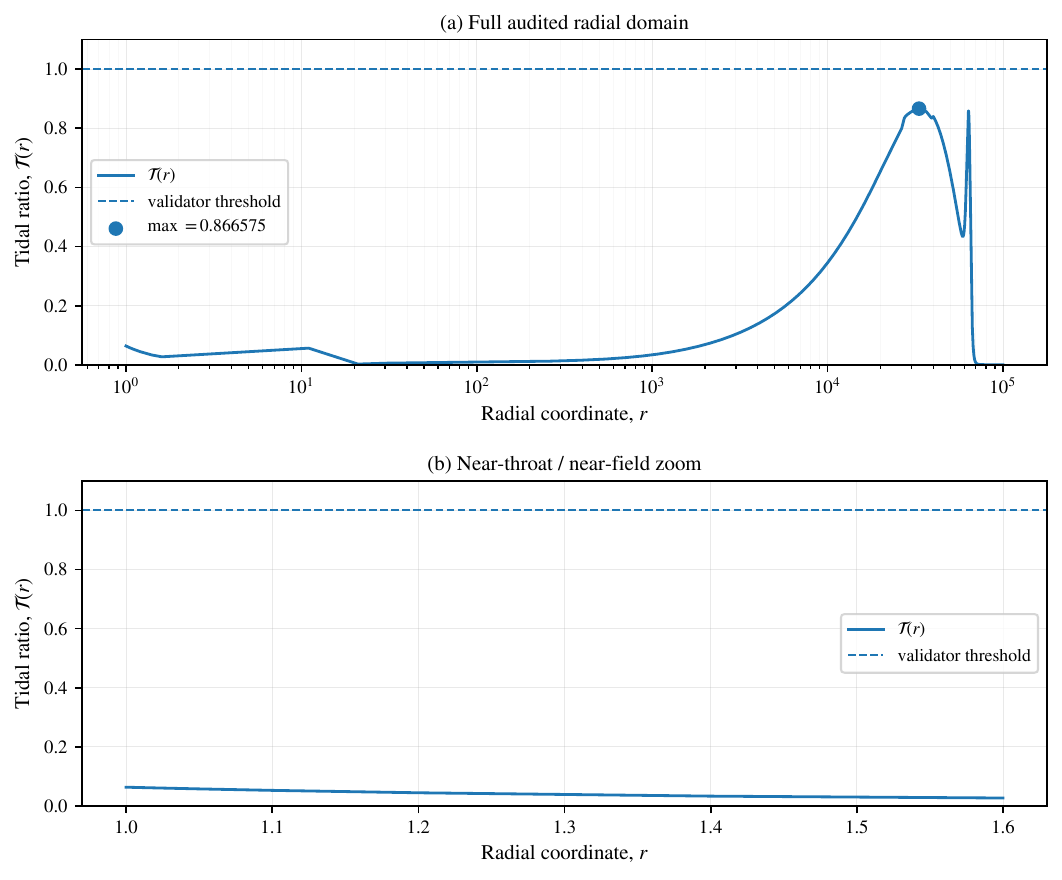}
\caption{
Archived validator-normalized tidal-curvature profile for the
selected finite-domain geometry. The dimensionless quantity
$\mathcal T(r)$ is read directly from the frozen validation archive,
with $\mathcal T=1$ representing the validator reference
normalization. The profile satisfies
$\max_i\mathcal T(r_i)=0.8665754053323194<1$ on the audited grid,
with the maximum near $r\simeq3.3204\times10^4$.
The reference level is an internal numerical normalization and is
not a universal observer-dependent safe-traversal criterion.
}
\label{fig:tidal_ratio_6_3}
\end{figure}

\FloatBarrier

The result in Eq.~\eqref{eq:tidal_max_value_location_6_3}
therefore establishes only that the archived dimensionless tidal
profile remains below its validator reference normalization on the
audited finite grid. It does not constitute an observer-dependent
safe-traversal test.

A physical traversal analysis would additionally require a physical
throat scale, an explicitly specified traveller trajectory and
velocity, a corresponding boosted orthonormal frame, body dimensions
or separation vectors, and admissible proper-acceleration and
tidal-acceleration limits \cite{MorrisThorne1988,Visser1995}.

Together with the finite-redshift and local proper-distance analysis
of Sec.~\ref{sec:wormhole_geometry}, the archived tidal result
provides one additional finite-domain geometric diagnostic. It does
not promote the reconstructed object to a globally completed or
observer-certified traversable wormhole spacetime.

Taken together, the three diagnostics in this section form a
post-selection validation layer on the frozen reconstruction. They
test reconstructed-source algebraic margins, a finite-domain radial
integral, and normalized tidal curvature without altering the
underlying geometry or curvature-sector reconstruction. Their
agreement therefore strengthens the internal consistency of the
reported finite-domain object, while the physical claim boundaries
associated with source microphysics, affine-complete null-energy
averaging, and observer-dependent traversal remain unchanged.


\section{Results for the frozen inverse reconstruction}
\label{sec:results}

The results below consolidate the geometric, curvature-sector,
source-side, null-energy, tidal, and provenance diagnostics defined in
Secs.~3--6. Every numerical quantity is evaluated on the same frozen
reconstruction and the same common finite radial grid. No metric
profile, curvature function, reconstructed-source array, closure
channel, radial null-energy array, or tidal profile is modified or
refitted at the Results stage. The numerical evidence is therefore
reported as properties of one post-selection inverse-reconstruction
object rather than as independently optimized diagnostic outcomes.

\subsection{Prescribed geometry on the frozen finite domain}
\label{subsec:geometry_results}

The selected Morris--Thorne-type geometry is evaluated on
\begin{equation}
\mathcal D_r
=
\{r_i\}_{i=1}^{10019},
\qquad
r_i\in[1,10^5],
\qquad
r_0=1 .
\end{equation}

The throat condition is imposed analytically by the compactness
parameterization,
\begin{equation}
B(1)=1,
\qquad
b(r_0)=r_0 ,
\end{equation}
and the local flare-out condition follows from
\begin{equation}
b'(r_0)=1-\eta,
\qquad
1-b'(r_0)=\eta>0 .
\end{equation}
These properties are therefore not obtained by post-selection
numerical fitting.

For the exterior radial metric factor
\begin{equation}
H(r)=1-\frac{b(r)}{r},
\end{equation}
Sec.~\ref{sec:wormhole_geometry} establishes
$H(r)>0$ analytically for $r>r_0$ within the adopted
parameterization. The smallest positive exported finite-grid value,
\begin{equation}
\min\nolimits^{+}_{r>r_0}H(r)
=
2.220446\times10^{-16},
\end{equation}
is consequently treated only as a floating-point cross-check near
the throat, not as the basis of the exterior no-horizon statement.

The frozen redshift profile satisfies
\begin{equation}
\begin{aligned}
\Phi_{\min}&=-3.139897645936\times10^{-4},
\\
\Phi_{\max}&=9.920371941187\times10^{-2}.
\end{aligned}
\end{equation}
with
\begin{equation}
\begin{aligned}
\max_i|\Phi(r_i)|&=9.920371941187\times10^{-2},
\\
\min_i e^{2\Phi(r_i)}&=9.993722176087\times10^{-1}.
\end{aligned}
\end{equation}
Thus the temporal metric coefficient remains finite and strictly
nonzero throughout the audited grid.

Figure~\ref{fig:metric_functions} provides the principal profile
view, while the detailed throat, flare-out, radial-margin, redshift,
and grid diagnostics are retained in the Supplementary Material.
The result established by this part of the frozen-object audit is
geometric and finite-domain:
the adopted parameterization contains an exact throat, analytic
local flare-out, positive exterior $H(r)$, bounded redshift, and
finite local proper distance at the throat. It does not establish
an exterior matching or a globally completed dynamical spacetime.

\subsection{Derivative-consistent curvature-sector reconstruction}
\label{subsec:results_reconstructed_fr_profiles}

The curvature sector reported here is the final
derivative-consistent inverse-reconstruction object defined in Sec.~\ref{sec:fr_reconstruction}. It is evaluated
on the sampled trajectory
\begin{equation}
\mathcal T_R
=
\{(r_i,R_i):R_i=R(r_i),\ r_i\in\mathcal D_r\},
\end{equation}
whose finite sampled curvature interval is
\begin{equation}
-1.1317148307\times10^{-9}
\le
R
\le
1.9999999979 .
\end{equation}

The archived discrete trajectory is non-monotonic under the
trajectory audit used for the frozen object and contains one
detected turning point. No numerical inverse map $r=r(R)$ is used
anywhere in the reconstruction. Repeated radial samples of the same
curvature therefore map to the same analytic values of
$f(R)$, $f_R(R)$, and $f_{RR}(R)$.

The final analytic reconstruction satisfies
\begin{equation}
\frac{df}{dR}=f_R,
\qquad
\frac{df_R}{dR}=f_{RR}
\end{equation}
by construction. On the common $N=10019$ nodes, the canonical
curvature arrays give
\begin{equation}
\begin{aligned}
\min_i f_R(R_i)
&=
0.02500508638762292,
\\
\max_i f_R(R_i)
&=
405.9949227252392,
\\
\min_i f_{RR}(R_i)
&=
0.0678574418696333,
\\
\max_i f_{RR}(R_i)
&=
980.4302876966746 .
\end{aligned}
\end{equation}
Consequently,
\begin{equation}
f_R(R_i)>0,
\qquad
f_{RR}(R_i)>0
\end{equation}
at every audited node.

The broad derivative ranges are reported explicitly because they
are part of the frozen curvature object and are relevant to
conditioning. Their positivity is interpreted only as a local
finite-domain curvature-sector diagnostic. It is not a proof of
perturbative stability, global stability, uniqueness, or viability
outside the sampled curvature interval.

Figure~\ref{fig:fr_reconstruction} shows the corresponding
$R(r)$ trajectory and the canonical $f(R)$, $f_R(R)$, and
$f_{RR}(R)$ profiles. The radial pullback and the detailed
finite-domain curvature audit are provided in the Supplementary
Material.

Thus, the reported curvature-sector result is not merely agreement
among three sampled arrays: it is the evaluation of one analytic
derivative hierarchy on the sampled curvature trajectory.

\subsection{Post-selection source-side closure}
\label{subsec:results_residual_closure}

The closure result is evaluated as the post-selection validation
functional defined in Sec.~\ref{subsec:channelwise_field_equation_residuals}.
For each archived channel
\begin{equation}
c\in\{tt,rr,\theta\theta,\mathrm{trace}\},
\end{equation}
the stored closure difference is
\begin{equation}
\mathcal C_c(r_i)
=
Q^{\rm base}_c(r_i)
-
Q^{\rm src}_c(r_i).
\end{equation}

The global finite-domain closure magnitude is
\begin{equation}
\|\mathcal C\|_{\infty,\mathcal D_r}
=
5.324612936468367\times10^{-5}.
\end{equation}
The corresponding archived baseline scale is
\begin{equation}
\|Q^{\rm base}\|_{\infty,\mathcal D_r}
=
0.7447379557892153,
\end{equation}
giving the relative maximum-norm reduction
\begin{equation}
\eta_{\mathcal C}
\simeq
99.992850\%.
\end{equation}

The channel-wise closure maxima are
\begin{equation}
\begin{aligned}
\|\mathcal C_{tt}\|_\infty
&=
2.596728353210916\times10^{-5},
\\
\|\mathcal C_{rr}\|_\infty
&=
1.331156343029403\times10^{-5},
\\
\|\mathcal C_{\theta\theta}\|_\infty
&=
1.331159735419005\times10^{-5},
\\
\|\mathcal C_{\rm trace}\|_\infty
&=
5.324612936468367\times10^{-5}.
\end{aligned}
\end{equation}
The trace channel therefore sets the reported global maximum.

Direct archive-level subtraction of the frozen baseline and
source-side arrays reproduces the stored closure columns to
roundoff-level precision,
\begin{equation}
\Delta_{\rm rec}^{\max}
\le
8.33\times10^{-17}.
\end{equation}

Figure~\ref{fig:residual_absmax_reduction_5_2} gives the compact
maximum-norm comparison, while the full four-channel radial profiles
are retained in the Supplementary Material.

These numbers establish internal numerical closure of the archived
baseline and reconstructed source-side channel representation.
They do not constitute an independent covariant solution test of
Eq.~\eqref{eq:reference_nmc_field_equation} for a uniquely
specified set
$\{\lambda,h(R),\mathcal L_m\}$.

\subsection{Post-selection reconstructed-source and radial null-energy results}
\label{subsec:results_energy_condition_anec}

The post-selection audit of the frozen reconstructed-source arrays
shows that $\rho$, $p_r$, and $p_t$ retain positive algebraic
energy-condition combinations throughout the audited finite grid.
The minimum values are
\begin{equation}
\begin{aligned}
\min_i\rho_i
&=
6.000000\times10^{-2},
\\
\min_i(\rho_i+p_{r,i})
&=
8.000000\times10^{-2},
\\
\min_i(\rho_i+p_{t,i})
&=
8.000000\times10^{-2},
\\
\min_i(\rho_i+p_{r,i}+2p_{t,i})
&=
1.195736\times10^{-1},
\\
\min_i(\rho_i-|p_{r,i}|)
&=
4.000000\times10^{-2},
\\
\min_i(\rho_i-|p_{t,i}|)
&=
4.000000\times10^{-2}.
\end{aligned}
\end{equation}

Thus, the reconstructed triplet satisfies the algebraic inequalities
associated with the WEC, radial NEC, tangential NEC, SEC, radial
DEC, and tangential DEC on $\mathcal D_r$.
Figure~\ref{fig:energy_condition_margins_6_1} shows the corresponding
radial profiles.

This result characterizes the reconstructed source representation
only. It does not identify an ordinary-matter Lagrangian, and it
does not establish that all possible exotic support has been
eliminated from a fully specified curvature--matter-coupled theory.

The radial null-energy combination
\begin{equation}
\mathcal N_r(r_i)
=
\rho_i+p_{r,i}
\end{equation}
remains positive at every audited node, with
\begin{equation}
\min_i\mathcal N_r(r_i)
=
8.000000\times10^{-2}.
\end{equation}
Using the archival notation $I_{\rm ANEC}$ for the cumulative
coordinate-radial integral,
\begin{equation}
I_{\rm ANEC}(r)
=
\int_{r_0}^{r}
[\rho(x)+p_r(x)]\,dx ,
\end{equation}
the frozen reconstruction gives
\begin{equation}
\begin{aligned}
I_{\rm ANEC}(10^5)
&=
8000.52020616,
\\
I_{\rm ANEC}(1.4)
&=
0.3522719805583363 .
\end{aligned}
\end{equation}
Both values are reproduced by direct coordinate-trapezoidal
integration of the same frozen $\rho+p_r$ array.

Figure~\ref{fig:cumulative_anec_6_2} shows the cumulative radial
profile. These positive quantities are finite-domain
coordinate-radial diagnostics. They are not affine-complete
averaged-null-energy integrals over complete null geodesics.

\subsection{Post-selection validator-normalized tidal-curvature result}
\label{subsec:results_tidal_gate}

The final tidal result is evaluated directly from the archived
dimensionless validator-normalized profile $\mathcal T(r)$ retained
in the frozen object and defined in Sec.~\ref{subsec:tidal_force_traversability_audit}. On the common
frozen grid,
\begin{equation}
\begin{aligned}
\max_i\mathcal T(r_i)
&=
0.8665754053323194,
\\
r_{\mathcal T,\max}
&\simeq
33203.988 .
\end{aligned}
\end{equation}

The complete archived profile therefore remains below the validator
reference normalization $\mathcal T=1$ on the audited finite domain.
The maximum occurs in the extended radial region rather than at the
throat.

Figure~\ref{fig:tidal_ratio_6_3} provides the corresponding radial
profile. The result is interpreted only as a property of the
archived normalized tidal diagnostic. The present reconstruction
does not independently restore a dimensional reference-curvature
scale from which a universal physical safety threshold could be
inferred.

Accordingly,
$\max\mathcal T<1$ is not an observer-dependent safe-traversal
statement. A physical traversal analysis would additionally require
a physical throat scale, traveller trajectory and velocity, boosted
orthonormal frame, body dimensions or separation vectors, and
admissible proper- and tidal-acceleration bounds.

\subsection{Consolidated frozen-object audit}
\label{subsec:results_summary_passed_validation_gates}

Table~\ref{tab:results_passed_validation_gates_7_6} consolidates the finite-domain evidence and the corresponding
claim boundaries for the frozen reconstruction. All entries refer to
the same $N=10019$ radial nodes and to quantities evaluated after the
reconstruction was fixed.

\begin{table*}[!htbp]
\centering
\scriptsize
\setlength{\tabcolsep}{3.6pt}
\renewcommand{\arraystretch}{1.10}
\caption{
Finite-domain result summary and associated claim boundaries for
the frozen reconstruction. The entries report what is established
by the archived numerical object and, separately, what is not
implied by that diagnostic.
}
\label{tab:results_passed_validation_gates_7_6}
\begin{tabularx}{\textwidth}
{@{}p{0.15\textwidth}p{0.24\textwidth}X p{0.25\textwidth}@{}}
\hline
Diagnostic
&
Numerical evidence
&
Supported statement
&
Not established
\\
\hline

Geometry
&
$r_0=1$; exact throat; $b'(r_0)<1$;
bounded $\Phi$; exterior $H>0$
&
The adopted Morris--Thorne parameterization satisfies the stated
geometric conditions on the audited finite domain.
&
Exterior matching, global continuation, or dynamical stability.
\\

Curvature sector
&
$\min f_R=0.0250051$;
$\min f_{RR}=0.0678574$;
$\max f_{RR}=980.4303$
&
The reconstructed curvature branch is derivative-consistent and
retains $f_R>0$ and $f_{RR}>0$ on the sampled trajectory.
&
Global $f(R)$ stability, uniqueness, or validity outside
$\mathcal I_R$.
\\

Archived source-side closure
&
$\|\mathcal C\|_\infty
=5.3246\times10^{-5}$;
$\eta_{\mathcal C}=99.992850\%$
&
The archived baseline and reconstructed source-side channels close
numerically under the frozen reconstruction.
&
Independent covariant solution of a uniquely specified
curvature--matter-coupled theory.
\\

Reconstructed source
&
All six reported pointwise minima are positive.
&
The reconstructed source triplet satisfies the corresponding
algebraic energy-condition inequalities on $\mathcal D_r$.
&
A unique microscopic matter model or elimination of all possible
exotic support.
\\

Coordinate-radial null energy
&
$8000.5202$ on $[1,10^5]$;
$0.352272$ on $[1,1.4]$
&
The cumulative coordinate-radial integrals are positive on both
reported finite intervals.
&
Affine-complete ANEC along arbitrary complete null geodesics.
\\

Normalized tidal profile
&
$\max\mathcal T=0.866575<1$
&
The archived dimensionless tidal profile remains below the
validator reference normalization.
&
Observer-dependent safe traversal or a universal human-safety
threshold.
\\

Internal recomputation
&
$\Delta_{\rm rec}^{\max}\le8.33\times10^{-17}$
&
The stored closure columns are arithmetically consistent with direct
subtraction of the corresponding frozen input arrays.
&
Independent third-party reproduction from separately obtained
inputs.
\\

\hline
\end{tabularx}
\end{table*}

\FloatBarrier

Taken together, the results establish one finite-domain
inverse-reconstruction object in which an analytically constrained
Morris--Thorne-type geometry, a derivative-consistent reconstructed
$f(R)$ sector, reconstructed source variables, coordinate-radial
null-energy diagnostics, archived source-side closure, a
validator-normalized tidal-curvature profile, and internal
recomputation checks are evaluated without result-level refitting.

The combined evidence establishes structural admissibility and
internal numerical consistency at the level explicitly defined in
Secs.~2--6. It is deliberately not promoted to a complete solution of
a uniquely specified nonminimally coupled $f(R)$ theory. The archive
does not determine a unique microscopic matter Lagrangian, coupling
strength, or coupling function, and the closure diagnostic is not an
independent covariant re-solution of such a model. Global existence,
uniqueness, dynamical stability, exterior matching, affine-complete
null-energy analysis, observer-dependent traversal, and independent
third-party reproduction therefore remain outside the result claimed
here.


\section{Discussion: inverse-problem interpretation and physical claim boundary}
\label{sec:discussion}

\subsection{Methodological interpretation of the finite-domain reconstruction}
\label{subsec:discussion_interpretation}

The principal result of this study is methodological rather than a
claim of a uniquely specified microscopic wormhole-support mechanism.
The calculation demonstrates that a geometry-conditioned functional
inverse reconstruction can be restricted to a structurally admissible
class before the final diagnostics are evaluated. For the present
Morris--Thorne test case, one geometry is reconstructed and audited on
the common frozen finite domain
\[
\mathcal D_r=\{r_i\}_{i=1}^{10019},\qquad
r_i\in[1,10^5],\qquad r_0=1,
\]
with no result-level refitting of the metric, curvature sector,
reconstructed source variables, closure channels, radial null-energy
data, or tidal profile.

Several ingredients of the construction are stronger than a
profile-only numerical fit. The throat and local flare-out
properties follow analytically from the metric parameterization,
the redshift sector remains bounded, and the exterior radial metric
factor is positive for $r>r_0$ within the adopted geometric ansatz.
The reconstructed curvature sector is derivative-consistent by
construction because $f_{RR}(R)$ is generated first and
$f_R(R)$ and $f(R)$ are obtained by integration. On the sampled
curvature trajectory, the final frozen branch satisfies
$f_R>0$ and $f_{RR}>0$ at every audited node
\cite{SotiriouFaraoni2010,DeFeliceTsujikawa2010}.

The remaining numerical diagnostics must be interpreted at the same
level of specificity. The quantities $\rho$, $p_r$, and $p_t$ are
reconstructed source variables. Their reported pointwise
energy-condition combinations remain positive on the finite grid,
and the corresponding coordinate-radial null-energy integrals are
positive over both the full audited interval and the selected
near-throat band. The archived tidal profile remains below its
validator reference normalization, while the archived source-side
closure difference remains small relative to the corresponding
baseline-channel scale. These results are mutually consistent
properties of one frozen numerical object.

From the inverse-problem perspective, the nontrivial feature is that
geometric admissibility, derivative consistency, source-side
diagnostics, tidal information, and numerical closure are evaluated
on one fixed reconstruction rather than on separately adjusted
objects. Nevertheless, none of these statements upgrades
the reconstruction to a complete covariant solution of a uniquely
specified curvature--matter-coupled theory. That stronger
interpretation would require physical information that is not
uniquely reconstructed here.

\subsection{Source-side closure and limits of exoticity inference}
\label{subsec:discussion_source_closure_exoticity}

The separation between the phenomenological reconstructed-source
representation and a uniquely specified microscopic theory is also
important for identifiability. The frozen source-side representation
does not contain enough information to identify a unique decomposition
of throat support into ``matter'', ``curvature'', and ``nonminimal
coupling'' contributions. Such a decomposition would require the
candidate-specific microscopic theory, including the matter
Lagrangian and coupling structure that are not reconstructed here.
The present inverse problem is therefore deliberately underdetermined
with respect to that microscopic decomposition, and no stronger
identification claim is made.

Accordingly, positivity of the algebraic combinations constructed
from $\rho$, $p_r$, and $p_t$ does not establish that an ordinary
microscopic matter model has been identified. It also does not prove
that all possible exotic support has disappeared. It establishes
only that the archived reconstructed-source representation satisfies
the reported pointwise inequalities on the audited finite domain.

The same caution applies to the closure result. The quantities
$Q_c^{\rm base}$ and $Q_c^{\rm src}$ are archived numerical
channels, and
\[
\mathcal C_c
=
Q_c^{\rm base}-Q_c^{\rm src}
\]
measures their source-side closure within the frozen numerical
representation. The small value of
$\|\mathcal C\|_{\infty,\mathcal D_r}$ and the roundoff-level
recomputation mismatch demonstrate internal arithmetic consistency.
They do not constitute an independent covariant re-solution of the
reference field equation for a unique
$\{\lambda,h(R),\mathcal L_m\}$.

\subsection{Finite-domain truncation and exterior completion}
\label{subsec:discussion_exterior_completion}

The finite-domain formulation is a deliberate part of the present
inverse reconstruction rather than an implicit claim of global
completion. No exterior matching is performed. Although the adopted
metric parameterization has a formally decreasing compactness profile
and a redshift function tending toward zero, the reconstructed
curvature and source sectors are validated only on their audited
finite domains. A formal asymptotic property of the metric ansatz
therefore does not determine a globally completed solution
\cite{MorrisThorne1988,Visser1995}.

A unique shell-free exterior problem cannot yet be posed from the
present archive alone. First, the reconstructed $f(R)$ sector is
validated only on the sampled curvature interval generated by the
finite-domain geometry. Second, no unique microscopic
$\mathcal L_m$, coupling strength $\lambda$, or coupling function
$h(R)$ has been reconstructed for continuation outside the audited
region. Third, the dimensionless throat normalization has not been
assigned a physical value. Choosing an exterior geometry and source
continuation before these ingredients are fixed would therefore add
new physical assumptions that are not determined by the present
inverse reconstruction.

A future exterior-completion problem would therefore constitute a new
inverse/continuation layer with additional physical assumptions. At
minimum, it would require specification of a matching radius and
exterior geometry, continuation or termination of the source sector,
selection of a candidate-specific gravitational and coupling model,
restoration of a physical throat scale, enforcement of the relevant
metric, extrinsic-geometric, and theory-dependent curvature matching
conditions, and re-evaluation of the full diagnostic hierarchy on the
completed spacetime.

The present paper therefore makes the narrower statement that the
finite-domain reconstruction is internally auditable; it does not
claim that a satisfactory exterior completion has been established.

\subsection{Relation to inverse problems, modified gravity, and
physics-informed reconstruction}
\label{subsec:discussion_relation_literature}

From the inverse-problem viewpoint, the main methodological contribution
is not the particular numerical values obtained for this wormhole
geometry, but the restriction of the reconstruction to a function
class that preserves exact differential identities before
post-selection validation. The hierarchy
\[
f_{RR}(R)\longrightarrow f_R(R)\longrightarrow f(R)
\]
reduces one source of structural ambiguity that would arise if the
three quantities were reconstructed independently. Likewise, direct
evaluation on the sampled trajectory $R(r)$ avoids introducing a
global inverse $r(R)$ when that map is not single-valued. These two
features concern the formulation of the reconstruction itself and are
logically distinct from the subsequent wormhole-specific physical
diagnostics.

Relative to the $f(R)$ wormhole literature, the principal
methodological distinction is the treatment of the curvature sector.
Many analytical and numerical studies examine how modified curvature
terms alter the throat-support and energy-condition balance
\cite{DeBenedictisHorvat2011,PavlovicSossich2014,
SamantaGodani2019,GolchinMehdizadeh2019,
GhoshMitraChakraborty2021}.
Here, however, $f(R)$, $f_R(R)$, and $f_{RR}(R)$ are not fitted as
three independent dense arrays. Their derivative consistency is
built into the reconstruction before the later diagnostics are
evaluated. The conditions $f_R>0$ and $f_{RR}>0$ are therefore
properties of one analytic curvature object, although they remain
local finite-domain diagnostics rather than a stability theorem
\cite{SotiriouFaraoni2010,DeFeliceTsujikawa2010}.

The nonminimal curvature--matter-coupling literature provides a
useful theoretical context because it shows that the separation
between explicit matter variables and curvature-dependent support is
model dependent
\cite{BertolamiEtAl2008,HarkoLobo2010,
GarciaLobo2010a,GarciaLobo2010b}.
The present study should not, however, be interpreted as another
fully specified model in that class. The reference coupling equations
motivate the claim boundary, whereas the numerical object actually
audited here consists of the reconstructed geometry, the
derivative-consistent $f(R)$ sector, phenomenological source variables,
and archived source-side closure channels.

The relation to the traversability literature is similarly limited.
A finite redshift function, finite local proper distance, and a
bounded tidal diagnostic provide useful geometric information, but
physical traversal requires an observer-dependent calculation
involving physical scale, trajectory, velocity, orthonormal-frame
curvature, body dimensions, and acceleration tolerances
\cite{MorrisThorne1988,Visser1995,
KuhfittigGladney2017,GarattiniChannuie2023}.
The validator-normalized quantity $\mathcal T(r)$ is therefore
retained as a numerical tidal-curvature diagnostic, not as a
human-safety criterion.

Finally, the role of physics-informed learning in the present work
is methodological rather than evidential by itself. Physics-informed
neural methods are capable of incorporating differential constraints,
but their known sensitivity to loss weighting, gradient imbalance,
sampling, and optimization makes a small training residual
insufficient as a physical proof
\cite{RaissiEtAl2019,WangTengPerdikaris2020,
CuomoEtAl2022,Basir2022}.
The frozen-object policy, analytic geometric constraints,
derivative-consistent curvature reconstruction, explicit diagnostic
audits, and archive-level recomputation therefore separate candidate
construction from the evidential layer. In this sense, machine
learning supplies a flexible representation and search mechanism,
whereas the strongest structural claims are carried by analytic
constraints and post-selection audits rather than by training loss
alone.

The transferable element of the framework is consequently structural
rather than geometry specific. The numerical values reported in this
paper---including the sampled curvature interval, energy-condition
margins, closure magnitude, and tidal profile---belong only to the
selected Morris--Thorne reconstruction. By contrast, the strategy of
embedding hard admissibility constraints in the representation,
generating derivative-linked functions through an integration
hierarchy, avoiding unnecessary inversion of a non-monotonic sampled
map, and freezing the selected object before validation can in
principle be applied to other functional inverse problems with analogous
structural relations. Demonstrating such transfer beyond the present
test case is a subject for future work rather than a result claimed
here.

\subsection{Limitations, transferability, and next validation steps}
\label{subsec:discussion_limitations_open_problems}

The principal limitations can be grouped into five distinct levels
rather than restating the same finite-domain qualification for each
diagnostic.

First, the reconstruction is not globally completed. No exterior
matching has been performed, and the reconstructed curvature sector
has not been validated outside the sampled curvature interval.
Cosmological viability, weak-field compatibility, and global
scalar-sector stability therefore remain untested
\cite{SotiriouFaraoni2010,DeFeliceTsujikawa2010}.

Second, the study is static. The positive curvature-sector sign
diagnostics and small archived source-side closure do not test linear
perturbations, nonlinear time evolution, collapse or scattering
response, or sensitivity to perturbations of the reconstructed source
variables. A dynamical-stability analysis is a separate requirement.

Third, the reconstructed source representation is phenomenological.
No unique microscopic matter Lagrangian, equation of state, source
conservation model, coupling strength, or curvature--matter coupling
function is inferred. Consequently, the present source inequalities
cannot settle the microscopic exoticity question.

Fourth, neither the radial null-energy nor tidal calculations are
complete physical traversability tests. The reported radial integral
is coordinate based and finite-domain rather than affine-complete,
while the tidal profile is validator normalized rather than
observer calibrated. A physical traversal study would require a
globally completed metric, a physical throat scale, complete null
geodesics, and explicitly specified observer trajectories and
acceleration limits.

Fifth, robustness, transferability, and independent reproducibility
remain separate from internal arithmetic consistency. The present
result concerns one selected frozen reconstruction, not a statistical
ensemble over architectures, initialization seeds, parameterizations,
sampling choices, or optimization settings. It also does not yet
demonstrate the same reconstruction strategy on a second physical
forward model. Robustness and transferability therefore remain
distinct future validation questions.

Robustness would require controlled perturbations of the metric
representation, curvature generator, sampling, tolerances, and
reconstruction settings followed by re-evaluation of the complete
diagnostic set. Likewise, the roundoff-level internal recomputation
does not constitute independent third-party reproduction. The
versioned public release of the frozen arrays, scripts, software
environment, and checksums provides the appropriate basis for such a
test
\cite{Peng2011,WilsonEtAl2014,WilkinsonEtAl2016}.

These limitations define the next scientific steps without changing
the status of the present result. The current contribution is an
auditable finite-domain functional inverse-reconstruction benchmark.
Its methodological extension would require robustness and
cross-problem transfer studies, whereas promotion to a globally
completed physical wormhole model would additionally require exterior
matching, candidate-specific microphysical modelling, dynamical
stability, affine-complete null-energy analysis, and
observer-dependent traversal calculations. Independent external
reproduction remains a further evidential layer enabled, but not
claimed, by the public archive.


\section{Conclusions}
\label{sec:conclusions}

This study establishes a finite-domain constrained functional
inverse-reconstruction benchmark using a Morris--Thorne geometry as a
physically demanding test case. The central methodological feature is
that structural admissibility is imposed before post-selection
validation: the metric parameterization embeds the required local
geometric conditions, while the reconstructed $f(R)$ sector is
generated through a derivative-consistent hierarchy rather than by
independently fitting $f$, $f_R$, and $f_{RR}$. The selected numerical
object is then frozen before source-side, null-energy, tidal,
closure, and provenance diagnostics are evaluated. The result is
therefore an auditable inverse reconstruction rather than a collection
of separately optimized diagnostic profiles.

The selected geometry is evaluated on
$N=10019$ radial nodes over $r\in[1,10^5]$ in throat-radius units.
The throat and local flare-out properties follow analytically from
the metric parameterization, the redshift sector remains bounded,
and the exterior radial factor satisfies $H(r)>0$ for $r>r_0$
within the adopted ansatz. The curvature sector is generated from a
single positive analytic $f_{RR}(R)$ representation and integrated
to obtain $f_R(R)$ and $f(R)$. On the sampled curvature trajectory,
\begin{equation}
\begin{aligned}
\min_i f_R(R_i)
&=
0.02500508638762292,
\\
\min_i f_{RR}(R_i)
&=
0.0678574418696333,
\\
\max_i f_{RR}(R_i)
&=
980.4302876966746,
\end{aligned}
\end{equation}
so the final frozen curvature object retains
$f_R>0$ and $f_{RR}>0$ at every audited node. Together with the
integration construction, these values characterize one
derivative-consistent admissible branch on the sampled curvature
trajectory. They remain finite-domain curvature-sector diagnostics
and do not constitute a proof of uniqueness, global viability, or
perturbative stability.

The reconstructed source variables $\rho$, $p_r$, and $p_t$
satisfy all six reported algebraic energy-condition inequalities on
the same frozen grid. The corresponding coordinate-radial
null-energy integrals are positive over both the complete audited
interval and the selected near-throat band,
\begin{equation}
\begin{aligned}
I_{\rm ANEC}(10^5)&=8000.52020616,
\\
I_{\rm ANEC}(1.4)&=0.3522719805583363.
\end{aligned}
\end{equation}
Here $I_{\rm ANEC}$ is retained only as the archive notation for a
finite-domain coordinate-radial integral; these values are not
affine-complete averaged-null-energy integrals along complete null
geodesics.

The archived source-side closure satisfies
\begin{equation}
\begin{aligned}
\|\mathcal C\|_{\infty,\mathcal D_r}
&=5.324612936468367\times10^{-5},
\\
\eta_{\mathcal C}&\simeq99.992850\%.
\end{aligned}
\end{equation}
and direct subtraction of the corresponding frozen baseline and
source-side arrays reproduces the archived closure columns to
roundoff-level accuracy,
\begin{equation}
\Delta_{\rm rec}^{\max}
\le
8.33\times10^{-17}.
\end{equation}
These results establish internal numerical consistency of the
archived source-side representation. They are not an independent
covariant re-solution of the reference field equation for a unique
set $\{\lambda,h(R),\mathcal L_m\}$.

The archived dimensionless tidal-curvature profile likewise remains
below its validator reference normalization,
\begin{equation}
\max_i\mathcal T(r_i)
=
0.8665754053323194
<1.
\end{equation}
This result is a finite-domain normalized tidal diagnostic, not an
observer-dependent safe-traversal criterion.

Taken together, the results demonstrate that geometric admissibility,
derivative consistency, reconstructed-source diagnostics,
coordinate-radial null-energy information, archived source-side
closure, normalized tidal curvature, and arithmetic provenance can
be evaluated coherently on one fixed finite-domain inverse
reconstruction. The central numerical claim is therefore structural
and computational: the reported evidence belongs to one common
post-selection object whose derivative relations and archived
diagnostics can be audited without result-level refitting.

The physical interpretation remains deliberately narrower. The
reconstruction does not identify a unique microscopic matter
Lagrangian, nonminimal coupling strength, coupling function, or
candidate-specific $F_{\rm NMC}(r)$ whose complete covariant field
equations have been independently solved. The positive
reconstructed-source energy-condition margins therefore characterize
the phenomenological source representation used in this work and do
not establish the universal absence of exotic support in a fully
specified wormhole theory.

The transferable contribution of the present work is the reconstruction
strategy rather than the numerical values of this particular
Morris--Thorne example: hard geometric constraints are embedded in the
representation, derivative-linked unknown functions are generated
through an integration hierarchy, unnecessary inversion of a
non-monotonic sampled map is avoided, and the selected object is frozen
before validation. Transferability to other inverse problems has not
yet been demonstrated and remains a separate validation task.
Likewise, promotion of the present finite-domain benchmark to a
globally completed physical wormhole model would require exterior
matching, candidate-specific microphysical modelling, dynamical
stability analysis, affine-complete null-energy evaluation, and
observer-dependent traversal calculations. Independent third-party
reproduction remains a further evidential layer enabled by the public
archive but not claimed here.

\section*{Funding}

The author received no financial support for the research, authorship, and/or
publication of this article.

\section*{Conflict of interest}

The author declares no competing interests.

\section*{Author contribution}

Murat Metehan T\"urko\u{g}lu conceived the study, developed the reconstruction
and validation framework, performed the numerical analysis, interpreted the
results, and wrote the manuscript.

\section*{Data availability}

The complete frozen numerical archive supporting the finite-domain
results reported in this study is publicly available in the GitHub
repository

\begin{center}
\url{https://github.com/MeteMurat/finite-domain-wormhole-fr-reconstruction}.
\end{center}

The exact frozen numerical data snapshot used for the reported
results is identified by commit
\texttt{\seqsplit{7860ef85b9126c1a760bde5f4ac3ecb1d64cbea1}}.

The public archive contains the common $N=10019$ radial-grid data,
the derivative-consistent reconstructed $f(R)$, $f_R$, and $f_{RR}$
arrays, reconstructed source variables, archived source-side
closure channels, coordinate-radial null-energy data, the
validator-normalized tidal diagnostic, V22A2/V22C/V22D audit
artifacts, machine-readable inventories, and SHA-256 manifests.

File-level integrity information is provided in
\path{public_data/FROZEN_DATA_MANIFEST.json},
\path{public_data/FROZEN_DATA_INVENTORY.csv}, and
\path{public_data/SHA256SUMS.txt}.
No numerical data required to audit the finite-domain claims of this
manuscript are withheld pending acceptance.

\section*{Code availability}

The reconstruction, numerical-audit, diagnostic, recomputation, and
publication-figure scripts used in this study are publicly available
at

\begin{center}
\url{https://github.com/MeteMurat/finite-domain-wormhole-fr-reconstruction}.
\end{center}

The repository includes the V22 reconstruction and audit chain,
the principal geometry, energy-condition, coordinate-radial
null-energy, tidal, and reproducibility scripts, together with
publication-facing generators, a Python requirements file, a Conda
environment specification, execution instructions, and frozen-data
checksum verification tools.

For exact correspondence with the numerical archive reported in the
manuscript, the frozen-data publication commit is
\texttt{\seqsplit{7860ef85b9126c1a760bde5f4ac3ecb1d64cbea1}}.

Historical internal workflow labels retained in some frozen scripts
are preserved for provenance and are mapped to the final manuscript
terminology in
\path{HISTORICAL_INTERNAL_LABELS_AND_CLAIM_BOUNDARY.md}.
They do not supersede the terminology used in the final manuscript.

\section*{Reproducibility statement}

The numerical results reported in this manuscript are tied to one
versioned frozen reconstruction. The public reproduction package
contains the frozen numerical arrays used by the manuscript,
associated audit artifacts, analysis and figure-generation scripts,
software-environment specifications, execution instructions, and
SHA-256 manifests.

The frozen numerical archive can be verified directly on Windows
PowerShell using the checksum verifier supplied in
\path{public_data/VERIFY_SHA256.ps1}. The archived closure
columns can additionally be recomputed from the corresponding
baseline and reconstructed source-side arrays; the manuscript reports
a maximum arithmetic recomputation mismatch no larger than
$8.33\times10^{-17}$.

Two levels of verification are deliberately distinguished. Internal
recomputation establishes arithmetic consistency of the frozen
archive, whereas the public repository now permits external
researchers to download the same numerical object and independently
repeat the reported finite-domain audit. Public availability itself
is not presented as evidence that an independent third party has
already reproduced the full reconstruction.

The released material supports reproducibility at the level claimed
in this paper. It does not convert the finite-domain reconstruction
into an independently validated covariant solution of a uniquely
specified curvature--matter-coupled theory.

\bibliographystyle{unsrtnat}
\bibliography{references_EPJC}

\begin{thebibliography}{39}
\providecommand{\natexlab}[1]{#1}
\providecommand{\url}[1]{\texttt{#1}}
\expandafter\ifx\csname urlstyle\endcsname\relax
  \providecommand{\doi}[1]{doi: #1}\else
  \providecommand{\doi}{doi: \begingroup \urlstyle{rm}\Url}\fi

\bibitem[Tarantola(2005)]{Tarantola2005}
Albert Tarantola.
\newblock \emph{Inverse Problem Theory and Methods for Model Parameter
  Estimation}.
\newblock Society for Industrial and Applied Mathematics, Philadelphia, PA,
  2005.
\newblock \doi{10.1137/1.9780898717921}.

\bibitem[Engl et~al.(1996)Engl, Hanke, and Neubauer]{EnglHankeNeubauer1996}
Heinz~W. Engl, Martin Hanke, and Andreas Neubauer.
\newblock \emph{Regularization of Inverse Problems}, volume 375 of
  \emph{Mathematics and Its Applications}.
\newblock Kluwer Academic Publishers, Dordrecht, 1996.
\newblock ISBN 978-0-7923-4157-4.
\newblock \doi{10.1007/978-94-009-1740-8}.

\bibitem[Morris and Thorne(1988)]{MorrisThorne1988}
Michael~S. Morris and Kip~S. Thorne.
\newblock Wormholes in spacetime and their use for interstellar travel: A tool
  for teaching general relativity.
\newblock \emph{American Journal of Physics}, 56\penalty0 (5):\penalty0
  395--412, 1988.
\newblock \doi{10.1119/1.15620}.

\bibitem[Visser(1995)]{Visser1995}
Matt Visser.
\newblock \emph{Lorentzian Wormholes: From Einstein to Hawking}.
\newblock AIP Press, Woodbury, NY, 1995.

\bibitem[Ford and Roman(1996)]{FordRoman1995}
L.~H. Ford and Thomas~A. Roman.
\newblock Quantum field theory constrains traversable wormhole geometries.
\newblock \emph{Physical Review D}, 53\penalty0 (10):\penalty0 5496--5507,
  1996.
\newblock \doi{10.1103/PhysRevD.53.5496}.
\newblock URL \url{https://arxiv.org/abs/gr-qc/9510071}.

\bibitem[Bertolami et~al.(2008)Bertolami, Harko, Lobo, and
  Páramos]{BertolamiEtAl2008}
Orfeu Bertolami, Tiberiu Harko, Francisco S.~N. Lobo, and Jorge Páramos.
\newblock Non-minimal curvature-matter couplings in modified gravity, 2008.
\newblock URL \url{https://arxiv.org/abs/0811.2876}.

\bibitem[Harko and Lobo(2010)]{HarkoLobo2010}
Tiberiu Harko and Francisco S.~N. Lobo.
\newblock {$f(R,L_m)$} gravity.
\newblock \emph{The European Physical Journal C}, 70:\penalty0 373--379, 2010.
\newblock URL \url{https://arxiv.org/abs/1008.4193}.

\bibitem[Garc{\'i}a and Lobo(2010)]{GarciaLobo2010a}
Nadiezhda~Montelongo Garc{\'i}a and Francisco S.~N. Lobo.
\newblock Wormhole geometries supported by a nonminimal curvature-matter
  coupling.
\newblock \emph{Physical Review D}, 82:\penalty0 104018, 2010.
\newblock URL \url{https://arxiv.org/abs/1007.3040}.

\bibitem[Garc{\'i}a and Lobo(2011)]{GarciaLobo2010b}
Nadiezhda~Montelongo Garc{\'i}a and Francisco S.~N. Lobo.
\newblock Nonminimal curvature-matter coupled wormholes with matter satisfying
  the null energy condition.
\newblock \emph{Classical and Quantum Gravity}, 28:\penalty0 085018, 2011.
\newblock URL \url{https://arxiv.org/abs/1012.2443}.

\bibitem[DeBenedictis and Horvat(2012)]{DeBenedictisHorvat2011}
Andrew DeBenedictis and Dubravko Horvat.
\newblock On wormhole throats in {$f(R)$} gravity theory.
\newblock \emph{General Relativity and Gravitation}, 44:\penalty0 2711--2744,
  2012.
\newblock URL \url{https://arxiv.org/abs/1111.3704}.

\bibitem[Pavlovi{\'c} and Sossich(2015)]{PavlovicSossich2014}
Petar Pavlovi{\'c} and Marko Sossich.
\newblock Wormholes in viable {$f(R)$} modified theories of gravity and weak
  energy condition.
\newblock \emph{The European Physical Journal C}, 75:\penalty0 117, 2015.
\newblock URL \url{https://arxiv.org/abs/1406.2509}.

\bibitem[Samanta and Godani(2019)]{SamantaGodani2019}
G.~C. Samanta and Nisha Godani.
\newblock Validation of energy conditions in wormhole geometry within viable
  {$f(R)$} gravity.
\newblock \emph{The European Physical Journal C}, 79:\penalty0 623, 2019.
\newblock URL \url{https://arxiv.org/abs/1908.04406}.

\bibitem[Golchin and Mehdizadeh(2019)]{GolchinMehdizadeh2019}
Hanif Golchin and Mohammad~Reza Mehdizadeh.
\newblock Quasi-cosmological traversable wormholes in {$f(R)$} gravity.
\newblock \emph{The European Physical Journal C}, 79:\penalty0 777, 2019.
\newblock \doi{10.1140/epjc/s10052-019-7292-4}.

\bibitem[Ghosh et~al.(2021)Ghosh, Mitra, and
  Chakraborty]{GhoshMitraChakraborty2021}
Bikram Ghosh, Saugata Mitra, and Subenoy Chakraborty.
\newblock Some specific wormhole solutions in {$f(R)$}-modified gravity theory.
\newblock \emph{Modern Physics Letters A}, 36\penalty0 (5):\penalty0 2150024,
  2021.
\newblock \doi{10.1142/S0217732321500243}.

\bibitem[Venkatesha et~al.(2023)Venkatesha, Kavya, and
  Sahoo]{VenkateshaEtAl2023}
V.~Venkatesha, N.~S. Kavya, and P.~K. Sahoo.
\newblock Geometric structures of morris--thorne wormhole metric in
  {$f(R,L_m)$} gravity and energy conditions.
\newblock \emph{Physica Scripta}, 98\penalty0 (6):\penalty0 065020, 2023.

\bibitem[Rosa and Kull(2022)]{RosaKull2022}
Jo{\~a}o~Lu{\'i}s Rosa and Paul~Martin Kull.
\newblock Non-exotic traversable wormhole solutions in linear {$f(R,T)$}
  gravity.
\newblock \emph{The European Physical Journal C}, 82:\penalty0 1154, 2022.
\newblock \doi{10.1140/epjc/s10052-022-11135-w}.

\bibitem[Tangphati et~al.(2024)Tangphati, Banerjee, and
  Pradhan]{BanerjeeTangphatiPradhan2023}
Takol Tangphati, Ayan Banerjee, and Anirudh Pradhan.
\newblock Wormholes and energy conditions in {$f(R,T)$} gravity.
\newblock \emph{International Journal of Geometric Methods in Modern Physics},
  21\penalty0 (6):\penalty0 2450109, 2024.
\newblock \doi{10.1142/S0219887824501093}.

\bibitem[Rosa et~al.(2023)Rosa, Ganiyeva, and Lobo]{RosaGaniyevaLobo2023}
Jo{\~a}o~Lu{\'i}s Rosa, Nailya Ganiyeva, and Francisco S.~N. Lobo.
\newblock Non-exotic traversable wormholes in {$f(R,T_{ab}T^{ab})$} gravity.
\newblock \emph{The European Physical Journal C}, 83:\penalty0 1040, 2023.
\newblock \doi{10.1140/epjc/s10052-023-12232-0}.

\bibitem[Ganiyeva et~al.(2025)Ganiyeva, Rosa, and Lobo]{GaniyevaRosaLobo2025}
Gulmira Ganiyeva, Jo{\~a}o~Lu{\'i}s Rosa, and Francisco S.~N. Lobo.
\newblock Wormhole geometries in {$f(R,T^2)$} gravity satisfying the energy
  conditions, 2025.
\newblock URL \url{https://arxiv.org/abs/2502.19323}.

\bibitem[Raissi et~al.(2019)Raissi, Perdikaris, and
  Karniadakis]{RaissiEtAl2019}
M.~Raissi, P.~Perdikaris, and G.~E. Karniadakis.
\newblock Physics-informed neural networks: A deep learning framework for
  solving forward and inverse problems involving nonlinear partial differential
  equations.
\newblock \emph{Journal of Computational Physics}, 378:\penalty0 686--707,
  2019.
\newblock \doi{10.1016/j.jcp.2018.10.045}.
\newblock URL \url{https://doi.org/10.1016/j.jcp.2018.10.045}.

\bibitem[Li et~al.(2023)Li, Li, Pang, and Long-Gang]{LiLiPang2023}
Zhi-Han Li, Chen-Qi Li, Pang, and Long-Gang.
\newblock Solving einstein equations using deep learning, 2023.
\newblock URL \url{https://arxiv.org/abs/2309.07397}.

\bibitem[Luna et~al.(2024)Luna, Doneva, Font, Lien, and
  Yazadjiev]{LunaEtAl2024}
Raimon Luna, Daniela~D. Doneva, Jos{\'e}~A. Font, Jr-Hua Lien, and Stoytcho~S.
  Yazadjiev.
\newblock Quasinormal modes in modified gravity using physics-informed neural
  networks.
\newblock \emph{Physical Review D}, 109\penalty0 (12):\penalty0 124064, 2024.
\newblock \doi{10.1103/PhysRevD.109.124064}.

\bibitem[Ncube et~al.(2021)Ncube, Harmsen, and
  Cornell]{NcubeHarmsenCornell2021}
A.~M. Ncube, G.~Harmsen, and A.~S. Cornell.
\newblock Investigating a new approach to quasinormal modes: Physics-informed
  neural networks, 2021.
\newblock URL \url{https://arxiv.org/abs/2108.05867}.

\bibitem[Cornell et~al.(2022)Cornell, Ncube, and
  Harmsen]{CornellNcubeHarmsen2022}
A.~S. Cornell, S.~Ncube, and G.~Harmsen.
\newblock Using physics-informed neural networks to compute quasinormal modes,
  2022.
\newblock URL \url{https://arxiv.org/abs/2205.08284}.

\bibitem[Ait~Haddou(2023)]{AitHaddou2023}
Mohamed Ait~Haddou.
\newblock Quasi-normal modes of near-extremal black holes in drgt massive
  gravity using physics-informed neural networks (pinns), 2023.
\newblock URL \url{https://arxiv.org/abs/2303.02395}.

\bibitem[Wang et~al.(2021)Wang, Teng, and Perdikaris]{WangTengPerdikaris2020}
Sifan Wang, Yujun Teng, and Paris Perdikaris.
\newblock Understanding and mitigating gradient flow pathologies in
  physics-informed neural networks.
\newblock \emph{SIAM Journal on Scientific Computing}, 43\penalty0
  (5):\penalty0 A3055--A3081, 2021.
\newblock \doi{10.1137/20M1318043}.

\bibitem[Cuomo et~al.(2022)Cuomo, Di~Cola, Fabio, Rozza, Raissi, and
  Piccialli]{CuomoEtAl2022}
Salvatore Cuomo, Vincenzo~Schiano Di~Cola, Giampaolo Fabio, Gianluigi Rozza,
  Maziar Raissi, and Francesco Piccialli.
\newblock Scientific machine learning through physics-informed neural networks:
  Where we are and what's next.
\newblock 2022.
\newblock URL \url{https://arxiv.org/abs/2201.05624}.

\bibitem[Nalui and Bhattacharya(2025)]{NaluiBhattacharya2025}
Subhasis Nalui and Subhra Bhattacharya.
\newblock Designing wormholes in novel power-law {$f(R)$}: A mathematical
  approach with a linear equation of state.
\newblock \emph{The European Physical Journal C}, 85\penalty0 (10):\penalty0
  1124, 2025.

\bibitem[Sotiriou and Faraoni(2010)]{SotiriouFaraoni2010}
Thomas~P. Sotiriou and Valerio Faraoni.
\newblock {$f(R)$} theories of gravity.
\newblock \emph{Reviews of Modern Physics}, 82:\penalty0 451--497, 2010.
\newblock \doi{10.1103/RevModPhys.82.451}.
\newblock URL \url{https://arxiv.org/abs/0805.1726}.

\bibitem[De~Felice and Tsujikawa(2010)]{DeFeliceTsujikawa2010}
Antonio De~Felice and Shinji Tsujikawa.
\newblock {$f(R)$} theories.
\newblock \emph{Living Reviews in Relativity}, 13:\penalty0 3, 2010.
\newblock \doi{10.12942/lrr-2010-3}.
\newblock URL \url{https://arxiv.org/abs/1002.4928}.

\bibitem[Hawking and Ellis(1973)]{HawkingEllis1973}
S.~W. Hawking and G.~F.~R. Ellis.
\newblock \emph{The Large Scale Structure of Space-Time}.
\newblock Cambridge Monographs on Mathematical Physics. Cambridge University
  Press, Cambridge, 1973.
\newblock ISBN 9780521099066.
\newblock \doi{10.1017/CBO9780511524646}.

\bibitem[Wald(1984)]{Wald1984}
Robert~M. Wald.
\newblock \emph{General Relativity}.
\newblock University of Chicago Press, Chicago, 1984.
\newblock ISBN 9780226870335.
\newblock \doi{10.7208/chicago/9780226870373.001.0001}.

\bibitem[Kontou and Sanders(2020)]{KontouSanders2020}
Eleni-Alexandra Kontou and Ko~Sanders.
\newblock Energy conditions in general relativity and quantum field theory.
\newblock \emph{Classical and Quantum Gravity}, 37\penalty0 (19):\penalty0
  193001, 2020.
\newblock \doi{10.1088/1361-6382/ab8fcf}.
\newblock URL \url{https://arxiv.org/abs/2003.01815}.

\bibitem[Kuhfittig and Gladney(2017)]{KuhfittigGladney2017}
Peter K.~F. Kuhfittig and Vance~D. Gladney.
\newblock Noncommutative-geometry inspired charged wormholes with low tidal
  forces.
\newblock \emph{Journal of Applied Mathematics and Physics}, 5\penalty0
  (3):\penalty0 574--581, 2017.
\newblock \doi{10.4236/jamp.2017.53049}.

\bibitem[Garattini and Channuie(2024)]{GarattiniChannuie2023}
Remo Garattini and Phongpichit Channuie.
\newblock Traversable wormholes supported by holographic dark energy with a
  modified equation of state.
\newblock \emph{Nuclear Physics B}, 1005:\penalty0 116589, 2024.
\newblock \doi{10.1016/j.nuclphysb.2024.116589}.

\bibitem[Basir(2023)]{Basir2022}
Shamsulhaq Basir.
\newblock Investigating and mitigating failure modes in physics-informed neural
  networks {(PINNs)}.
\newblock \emph{Communications in Computational Physics}, 33:\penalty0
  1240--1269, 2023.
\newblock \doi{10.4208/cicp.OA-2022-0239}.
\newblock URL \url{https://arxiv.org/abs/2209.09988}.

\bibitem[Peng(2011)]{Peng2011}
Roger~D. Peng.
\newblock Reproducible research in computational science.
\newblock \emph{Science}, 334\penalty0 (6060):\penalty0 1226--1227, 2011.
\newblock \doi{10.1126/science.1213847}.

\bibitem[Wilson et~al.(2014)Wilson, Aruliah, Brown, Chue~Hong, Davis, Guy,
  Haddock, Huff, Mitchell, Plumbley, Waugh, White, and Wilson]{WilsonEtAl2014}
Greg Wilson, D.~A. Aruliah, C.~Titus Brown, Neil~P. Chue~Hong, Matt Davis,
  Richard~T. Guy, Steven H.~D. Haddock, Kathryn~D. Huff, Ian~M. Mitchell,
  Mark~D. Plumbley, Ben Waugh, Ethan~P. White, and Paul Wilson.
\newblock Best practices for scientific computing.
\newblock \emph{PLoS Biology}, 12\penalty0 (1):\penalty0 e1001745, 2014.
\newblock \doi{10.1371/journal.pbio.1001745}.

\bibitem[Wilkinson et~al.(2016)Wilkinson, Dumontier, Aalbersberg, Appleton,
  Axton, Baak, Blomberg, Boiten, da~Silva~Santos, Bourne,
  et~al.]{WilkinsonEtAl2016}
Mark~D. Wilkinson, Michel Dumontier, I.~Jsbrand~Jan Aalbersberg, Gabrielle
  Appleton, Myles Axton, Arie Baak, Niklas Blomberg, Jan-Willem Boiten,
  Luiz~Bonino da~Silva~Santos, Philip~E. Bourne, et~al.
\newblock The fair guiding principles for scientific data management and
  stewardship.
\newblock \emph{Scientific Data}, 3:\penalty0 160018, 2016.
\newblock \doi{10.1038/sdata.2016.18}.

\end{thebibliography}

\end{document}